\documentclass[sigconf,9pt]{acmart}
\usepackage[ruled,vlined]{algorithm2e}
\usepackage{caption}
\usepackage{subcaption}
\usepackage{adjustbox}
\usepackage{algpseudocode}
\usepackage{hyphenat}

\newcommand{\parahead}[1]{\vspace*{0.4ex plus 0.15ex minus 0.15ex}\noindent %
  {\bfseries #1.}}

\renewcommand{\paragraph}[1]{\vspace{2pt plus 0pt minus 2pt}\noindent{\bfseries #1}}

\newcommand{\systemname}{SiFall}
\newcommand{\systemnames}{SiFall's}
\AtBeginDocument{%
  \providecommand\BibTeX{{%
    \normalfont B\kern-0.5em{\scshape i\kern-0.25em b}\kern-0.8em\TeX}}}

\setcopyright{acmcopyright}
\copyrightyear{2022}
\acmYear{2022}
\acmDOI{10.1145/3560905.3568517}

\acmSubmissionID{75}

\begin{document}

\title{\systemname{}: Practical Online Fall Detection with RF Sensing}

\acmYear{2022}\copyrightyear{2022}
\setcopyright{acmcopyright}
\acmConference[SenSys '22]{ACM Conference on Embedded Networked Sensor Systems}{November 6--9, 2022}{Boston, MA, USA}
\acmBooktitle{ACM Conference on Embedded Networked Sensor Systems (SenSys '22), November 6--9, 2022, Boston, MA, USA}
\acmPrice{15.00}
\acmDOI{10.1145/3560905.3568517}
\acmISBN{978-1-4503-9886-2/22/11}

\author{Sijie Ji}
\email{sijie001@e.ntu.edu.sg}
\affiliation{%
   \institution{Nanyang Technological University}
   \streetaddress{50 Nanyang Ave}
   \city{Singapore}
   \country{Singapore}}

\author{Yaxiong Xie}
\email{yaxiongx@buffalo.edu}
\affiliation{%
   \institution{University at Buffalo}
   \streetaddress{321 Davis Hall}
   \city{Buffalo}
   \country{New York}}

\author{Mo Li}
\email{limo@ntu.edu.sg}
\affiliation{%
   \institution{Nanyang Technological University}
   \streetaddress{50 Nanyang Ave}
   \city{Singapore}
   \country{Singapore}}

\begin{abstract}
Falls are one of the leading causes of death in the elderly people aged 65 and above. In order to prevent death by sending prompt fall detection alarms, non-invasive radio-frequency (RF) based fall detection has attracted significant attention, 
due to its wide coverage and privacy preserving nature. 
Existing RF-based fall detection systems process fall as an activity classification problem 
and assume that human falls introduce reproducible patterns to the RF signals.
We, however, argue that the fall is essentially an accident, 
hence, its impact is uncontrollable and unforeseeable.  
We propose to solve the fall detection problem in a 
fundamentally different manner.
Instead of directly identifying the human falls which are difficult to quantify, we recognize the normal repeatable human activities and then identify the fall as abnormal activities out of the normal activity distribution.
We implement our idea and build a prototype based on commercial Wi-Fi.
We conduct extensive experiments with 16 human subjects. 
The experiment results show that our system can achieve high fall detection accuracy and adapt to different environments for real-time fall detection.

\end{abstract}




\maketitle

\section{Introduction}
\begin{figure}[t!]
  \centering
  \includegraphics[width=0.95\linewidth]{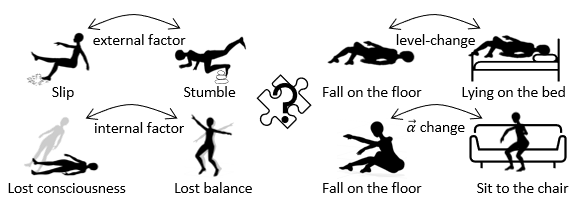}
  \caption{Diversity of falls.}
  \label{fig:mot1}
\end{figure}

Fall is an important global public health issue~\cite{world2008global}.
Every year there are approximately 37.3 million fall-related injuries that require medical attention and directly cost \$34 billion~\cite{burns2016direct}.
Clinical reports show that timely treatment ($<$1 hour) can prevent deaths from fatal falls~\cite{wilder1981dangerous}. 
Therefore, an effective fall detection system is necessary to facilitate timely treatment and 
benefit the current aging society where more and more elderly people are living alone~\cite{cheng2009global}.

Existing fall detection solutions can be classified into two categories: 
wearable-based solutions and device-free solutions.
Medical research has reported that wearable-based solutions 
do not work well in practice due to the burden of carrying 
and charging those devices from time to time~\cite{centers2010falls}. 
In contrast, device-free solutions including computer vision (CV) based, acoustic-based, and RF-based are more user-friendly.
Among them, the CV-based solutions cannot work under dim light conditions, occlusions and often compromises user privacy. 
The acoustic-based solutions are limited by its sensing range ($<$4.5m)~\cite{wang2020push} 
and possibly subject to restriction by ambient loudness ($<$40dB SPL)~\cite{li2020fm}. 
However, RF-based solutions are not constrained by the above and also cost-effective as they take the advantage of existing ubiquitous communication infrastructures such as WiFi APs.

Existing RF-based fall detection systems~\cite{wang2016wifall,wang2016rt,palipana2018falldefi,tian2018rf}
consider falls as a type of normal human activity and applies
traditional human activity recognition method to identify
the falls out of similar activities such as sitting, sleeping and
jumping.
Generally, the solution consists of off-line training and on-line inference. 
During the off-line training, the system builds up a model based on feature engineering~\cite{wang2016wifall,palipana2018falldefi} or machine learning~\cite{wang2016rt,tian2018rf}, to separate the falls from other human activities. 
The RF signals are collected for training purposes when the human being performs a set of pre-defined activities, such as falling, sitting and jumping. 
The system then applies the trained model to identify falls from the received signals.
\textit{All existing solutions implicitly assume that 
 human falls introduce reproducible patterns to the RF signals which can be captured by the trained model and used to differentiate the falls from other activities.}

In this paper, we revisit such a problem and argue that the signal patterns introduced by human falls are full of randomness and consequently hard to be fully captured by trained templates.
Our key intuition is that the human fall, by its nature, is an accident that is \textbf{unforeseeable} and the human reaction is highly \textbf{uncontrollable}, introducing highly dynamic disturbance to the wireless signals.
Specifically, as depicted in Figure~\ref{fig:mot1}, 
there are diverse causes of human falls, such as a stumble, a slip, loss of consciousness, loss of balance, a sudden fright, etc, which may result in randomness, e.g., a stumble or a slip
may result in displacement of the human body. In contrast, a person stays at the same place if he loses his consciousness. In addition, the free range of movement in the joints of human body brings in another level of randomness when the human being cannot properly control his behavior during the falls.
Extracting representative features of the human falls becomes impractical because of such uncontrolled randomness. 
Even collecting adequate data is challenging because one person can hardly repeat real and uncontrollable falls. 

\begin{figure*}[t]
    \begin{subfigure}[b]{0.60\textwidth}
		\includegraphics[width=\textwidth,height=6cm]{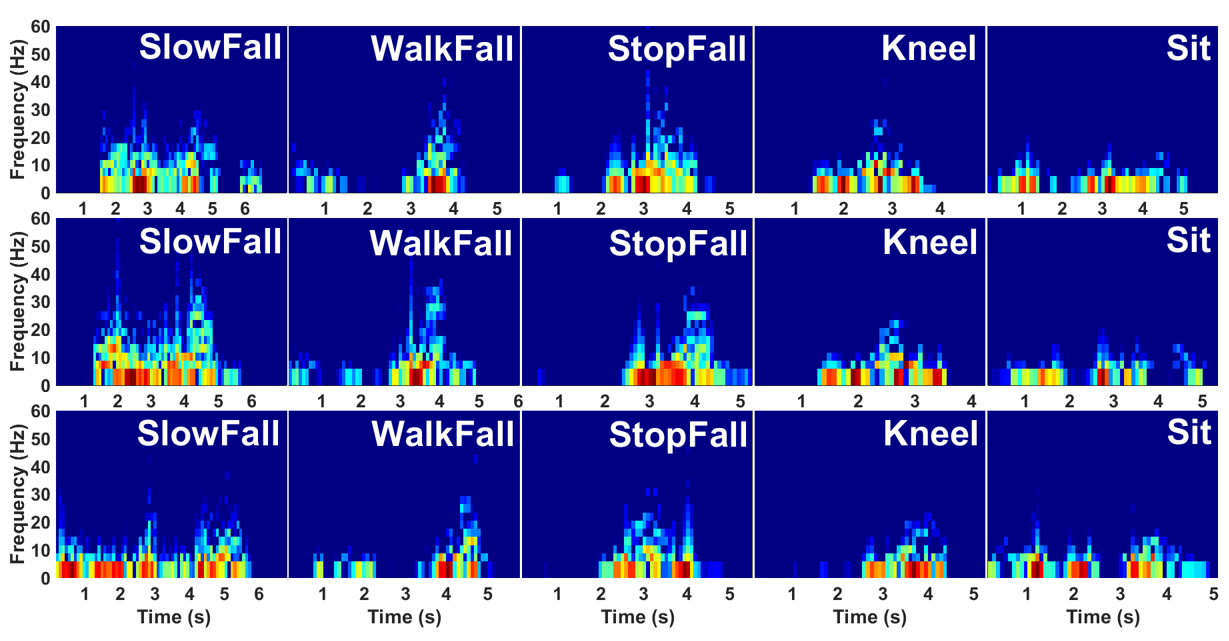}
		\centering
		\caption{STFT segments from the same testing subject \\
		\#Person 1}
		\label{fig:stftfl}
	\end{subfigure}
	\begin{subfigure}[b]{0.39\textwidth}
		\includegraphics[width=\textwidth,height=6cm]{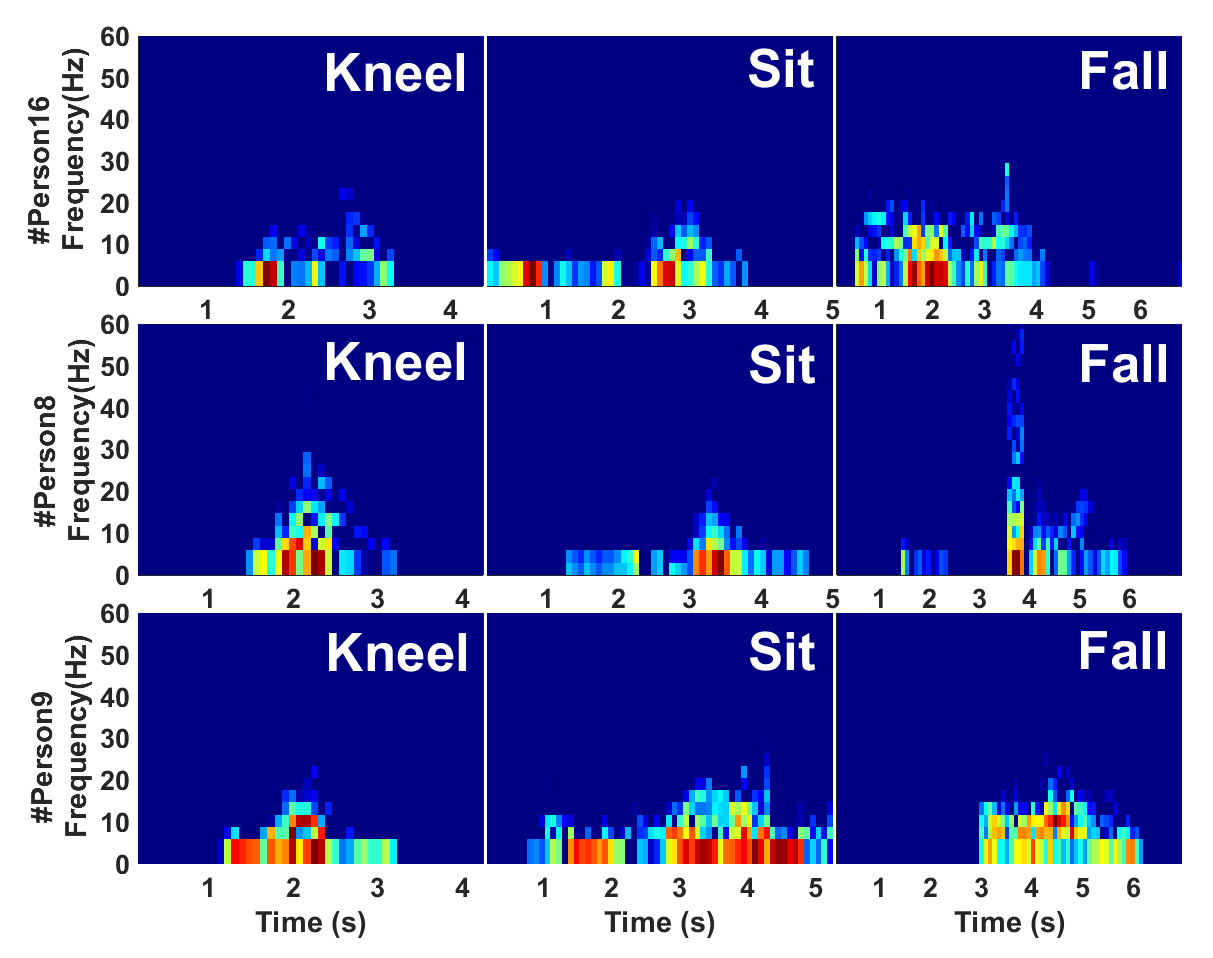}
		\centering
		\caption{STFT segments from different testing subjects \\    	\#Person 8, 9, and 16}
		\label{fig:stftfall}
	\end{subfigure}
    \caption{STFT segments across activities and testing subjects.}
    \label{fig:stft}
\end{figure*}

With the above observation, in this paper we handle the fall detection problem in a fundamentally different manner. Instead of seeking features to characterize the unforeseeable and uncontrollable human falls, we turn to solving an easier problem: recognizing normal repeatable human activities including but not limited to jumping, sitting, and walking.
We formulate fall detection as adaptive anomaly detection and identity an abnormal activity that cannot be classified as any of the known activities as a fall. Our hypothesis is that after an adequate time period of training, a self-supervised learning process will eventually perfect the model to differentiate uncontrolled falls from other repeated controlled human activities. To prune the search space and speed up the convergence of the model training, we apply analyzable signal processing to early filter out non-fall human activities with distinguishable signal features. Our observation suggests that falls change the status of the human body in a short period of time and thus introduce high frequency components to the signal variations.
We feed the identified suspicious fall-like activities to a deep neural network called FallNet to recognize the true falls. Specifically, the FallNet trains an auto-encoder~\cite{hinton2006reducing} to learn a compressed representation of normal fall-like activities. When used for inference, the auto-decoder is only able to accurately reconstruct the normal fall-like activities but not real human falls. The FallNet, therefore, identifies the activities that result in large reconstruction error as falls. After deployment, the FallNet is continuously updated using the freshly collected data in a self-supervised manner, so it evolves to adapt to the local propagation environment and the particular human subjects that the system monitors. We expect that the FallNet will eventually perfect its detection accuracy and false alarm rate over time.

To realize our idea, we implement a \textbf{S}elf-supervised \textbf{I}ncremental learning Fall detection system, \systemname{}.
To the best of our knowledge, \systemname{} is the first RF-based fall detection system that can work in real time for online fall detection on a daily basis
across different human, different environments and different types of activities. \systemname{} possesses the following three advantages: 
\begin{itemize}
    \item
    SiFall works with daily human activities in runtime - WiFi CSI samples are dynamically processed, segmented, and discriminated to detect ongoing "falls". 
    \item
    SiFall's self-learning process can adapt to the variation of human subjects, environment, and types of falls. The core anomaly detection model of SiFall evolves during its use;
    \item
    SiFall separates the signal processing from its machine learning model, which is designed to be lightweight and may easily be accommodated at the edge devices. 
\end{itemize}

The developed SiFall prototype has been comprehensively evaluated with a total amount of over 92 hours of test data collected from 16 human subjects of different ages and genders. During our experimental evaluation, SiFall is able to achieve 98.3\% accuracy in a real-world setting with extensive movements. During a continuous three-day adoption in a normal living environment, SiFall is able to detect 94.1\% falls with only one false alarm in the end.

\section{Challenges and Opportunities}
This section first discusses the challenges in developing a practical RF-based fall detection system and then presents the key observations and opportunities.

\subsection{Challenges}

\subsubsection{Fall Ambiguity.}
 There is no uniform quantitative definition of "fall" in medicine, biology, or physics. According to the World Health Organization (WHO), fall is a subjective term, which is measured by the level of discomfort in the human body after a person accidentally lies on the ground or other low level~\cite{world2008global}. As a result, it is hard to identify a quantified signal template to feature the "fall" when performing RF sensing. Besides, the orientation and the reflection surface of the human body may impact the reflected RF signal which leads to inter-activity similarities (e.g., falling v.s. lying down)~\cite{korany2019xmodal,adib2015capturing}. Other factors including deployment layout and individual difference may also contribute to the ambiguity in defining and quantifying the "fall" in the RF signal space.

\subsubsection{Data Scarcity}
Recent advances in deep learning allow learning powerful discriminative models from a number of representative samples~\cite{deng2009imagenet}, which may bypass the difficulty in defining precise signal templates of "falls". However, since the "falls" are high exceptional human activities that often occur uncontrollably, it is extremely difficult to obtain sufficient repeatable real-life data samples containing different types of falls, leading to a data scarcity issue. Most existing fall detection studies depend on learning from artificial fall samples collected from the laboratory environment and thus may have gaps in detecting real falls that take place in daily life. The lack of fall data may also result in class distribution skews where the learned model is biased towards the majority types of falls and may have poor predictive performance for other types of falls. As long as the types of falls are not sufficiently emulated, the learned model may be unreliable with poor generalizability.

\subsubsection{Unstructured Input Signal}
Human motions, even of the same type, may last for different durations of time, and as a result, the relevant RF signals are unstructured and of different lengths. The processing of variable-length input signals is very different from processing  fixed-length data samples in many machine learning models. Real-time processing of variable-length sequences is particularly difficult because data structurization techniques like sequence padding or dynamic template mapping can hardly be applied in real time~\cite{tan2019time}. In addition, real-time segmentation of the RF signals from consecutive activities is also challenging, the inaccuracy of which may lead to inconsistency of features in the machine learning model. Most existing fall detection solutions assume pre-defined fixed-length RF signal input.

\subsection{Opportunities}

While the practical challenges suggest extreme difficulties in learning the RF templates of human falls, we observe that there is an opportunity on the other hand to categorize human daily activities as they are usually repeatable and there exist plenty daily data samples for training a model to describe them. To showcase such an observation, Figure~\ref{fig:stftfall} visualizes the extracted WiFi signal features after short time Fourier transformation (STFT) across various human activities (details in \S 3.2). Figure~\ref{fig:stftfl} depicts the STFT segments collected from the same testing human subject and Figure~\ref{fig:stftfall} depicts those collected from three different human subjects. It is obvious to see that the daily human activities give very consistent STFT patterns, e.g., the kneeling and sitting patterns in Figure~\ref{fig:stftfl}. Even across different human subjects the patterns of the same daily activities remain consistent, e.g., the kneeling and sitting patterns in Figure~\ref{fig:stftfall}. The "falls" however appear highly varied and non-repeatable across the types, e.g., the "stop fall", "walk fall", and "slow fall" (details in \S4.2), as well as the testing human subjects. The above observations suggest that it is reliable to train a model to accurately describe the normal daily activities and as an opportunity to identify "falls" as abnormal outlier output from such a model. As there are plenty of daily activities to see when the system is deployed in reality, a self-supervised learning scheme may continuously perfect the trained model with improved accuracy in distinguishing the falls from normal daily activities.

\section{System Design}

A desired fall detection system should have the following characteristics: (i) it must work in real-time and detect falls with run-time data input; (ii) it must be able to evolve itself without involving human efforts to label the data samples;  (iii) it must adapt to environment and different users. In this section, we present the design of SiFall, a system that accommodates the above design considerations. We begin with the system overview followed by fall-like activity segmentation and the design of FallNet.

\subsection{Overview}
\begin{figure}[t]
  \centering
  \includegraphics[width=\linewidth]{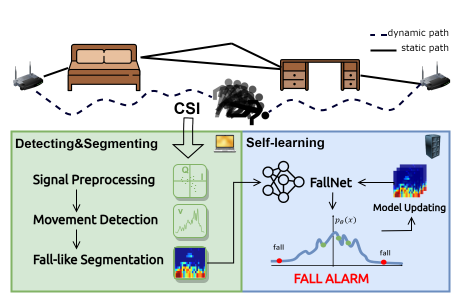}
  \caption{Overview of SiFall}
  \vspace{-0.2cm}
  \label{f:overview}
\end{figure}

 \systemname{} consists of a front-end to process RF signals and a back-end server to train the neural network model and detect the fall, as shown in Figure~\ref{f:overview}.

SiFall's front-end collects WiFi channel state information (CSI) measurements, denoises the CSI and extracts the dynamic component of the CSI to obtain an approximate RF-signal description of human movement. Finally, a lightweight algorithm is used to quantify the motion intensity and segment the RF signals accordingly (\S 3.2). 
In the end, SiFall applies short-time Fourier transformation (STFT) to derive the time-frequency spectrum of each piece of segmented RF signal clip 
and supplies the STFT spectrum to the back-end server for fall detection. The purpose of the front-end signal processing is two folded: to early rule out normal activities that possess clear daily activity features, and to present segmented RF signals with data cleansing. Typical daily human movements without high-frequency components are expected to be filtered out to narrow down the learning space of the back-end neural network model.

In the back-end server, a self-evolving deep neural network called FallNet takes the segmented RF signal as input
and identify the falls from the normal fall-like activities.
The FallNet is designed based on the auto-encoder framework to do the self-supervised learning where the encoder learns a nonlinear mapping from the unstructured RF-signal space to uniformed compact latent feature space and thus addresses challenge from the unstructured input signal. The decoder learns the mapping from the latent space back to the RF-signal space with the goal of reconstructing the original RF-signal as closely as possible. After training 
with a large number of repeated regular human activities, the FallNet establishes a Gaussian mixture distribution of normal
human activities in the latent space (\S 3.3) and thus is capable of accurately recognizing and recovering the RF-signal clips of normal activities. When deployed, the FallNet classifies
the fall-like RF-signal clips that can be well reconstructed as normal daily activities and those RF-signal clips that cannot be reconstructed as falls. The FallNet is continuously updated with RF-signal clips of repeatedly-appearing normal human activities fed from the front-end (\S 3.4).

\subsection{RF signal Segmentation}

\subsubsection{CSI Extraction and Denoising}
The received WiFi signal can be modeled as: 

\begin{equation}
    Y(f, t) = H(f,t) \times X(f, t)
\end{equation}

where $X(f, t)$ represents the signals carried at subcarrier frequency $f$ and time point $t$ and $H(f, t)$ denotes the CSI value at $f$. 
The CSI describes how the RF signals are transformed by the current wireless channel - 
the amplitude attenuation and phase rotation of different frequency components due to multipath reflection, diffraction, and scattering by objects in the environment. 
On top of that, RF chipset processing at WiFi transceivers may introduce additional distortion and noises~\cite{xie2018precise,xie2019md}. 
Therefore, we perform necessary data cleaning to eliminate the impact of the hardware imperfections. 

\begin{figure}[t]
  \centering
  \includegraphics[width=\linewidth]{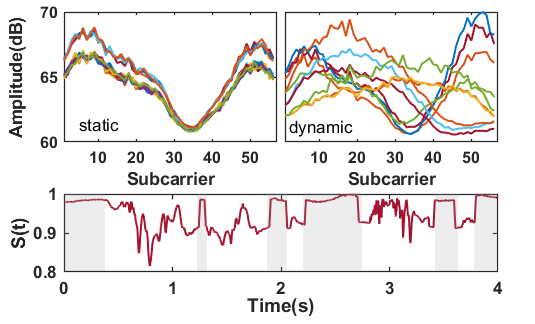}
  \caption{Static CSI Amplitude (upper left), Dynamic CSI Amplitude (upper right) and the Extract Channel Dynamic $S(t)$, gray is the ground truth static.}
 \vspace{-0.3cm}
  \label{fig:dy1}
\end{figure}

The CSI $H$ consists of a static part induced by ambient environment $H_{s}$ and 
a dynamic part related to human movement $H_{d}$. 
CSI is also subject to WiFi hardware distortion $H_{h}$.
Therefore, we model the overall CSI as: 
\setlength{\abovedisplayskip}{12pt}
\setlength{\belowdisplayskip}{12pt}
\setlength{\abovedisplayshortskip}{2pt}
\setlength{\belowdisplayshortskip}{2pt} 
\begin{equation} \label{eq1}
\begin{split}
H\left(f, t \right) & = \left( H_{s}\left(f, t \right) + H_{d}\left(f, t \right) \right) \cdot  H_{h}\left(f, t \right) \\
 & =  \left( H_{s}\left(f, t \right) + H_{d}\left(f, t \right) \right)  \cdot \varepsilon_{1}\left(t\right) e^{\varepsilon_{2}(f, t)+\varepsilon_{3}(t)+\varepsilon_{4}}
\end{split}
\end{equation}
where $\varepsilon_{1}(t)$ is the amplitude scaling caused by automatic gain control (AGC), $\varepsilon_{2}(f,t)$ represents the phase offset introduced by the combination of packet detection delay (PDD), 
sampling frequency offset (SFO) and sampling time offset (STO), 
$\varepsilon_{3}(t)$ is the phase offset caused by the carrier frequency offset (CFO),  
and $\varepsilon_{4}$ is the initial phase offset of the radio chains.
We utilize relatively clean CSI amplitude and mitigate the impact of noisy CSI phase by 
calculating the conjugate multiplication of CSI as $\hat{H}(f,t)$ for each subcarrier:
\begin{equation}
\hat{H}(f,t) \equiv H\left(f, t \right)\overline{H\left(f, t \right)}=\varepsilon_{1}^{2}\left(t\right) \left | H_{s}\left(f, t\right)+H_{d}\left(f, t\right) \right   |^{2},
\end{equation}
The resulting $\hat{H}(f,t)$ is still affected by the amplitude scaling $\varepsilon_{1}(t)$ that AGC introduces.
To visualize the impact of $\varepsilon_{1}(t)$, we collect CSI measurements from a static environment and calculate CSI amplitude across subcarriers in Figure~\ref{fig:dy1} (upper left),
from which we see that the CSI amplitude curves across subcarriers are similar but not identical.
The reason is that the amplitude scaling factor $\varepsilon_{1}(t)$ is time-varying but consistent across subcarriers.
We note that, because of the amplitude scaling factor, the CSI amplitude of a single subcarrier $\hat{H}(f,t)$ 
is time-varying even when the environment is static and thus cannot capture the dynamics introduced by the human motion.

\subsubsection{Capturing Channel Dynamics}
We use the variations of the CSI amplitude curve to
capture the channel dynamics introduced by human motion.
To illustrate the intuition, we plot the CSI amplitude when the human is moving in Figure~\ref{fig:dy1} (upper right),
from which we see that the shape of amplitude curve varies significantly in non-static environment.
We use \textit{cosine similarity} to quantify the similarity between consecutive CSI measurements:
\begin{equation}
 S(t_n) =     
 \frac{\langle \hat{H}(t_n), \hat{H}(t_{n-1}) \rangle}{|\hat{H}(t_n)||\hat{H}(t_{n-1})|}
\end{equation}
where $\hat{H}(t_n) = [\hat{H}(f_1, t_n), \cdots \hat{H}(f_M, t_n)]$ represents
the CSI amplitude vector of all $M$ subcarriers sampled at $n$-th time point.
We plot the calculated S(t) for CSI collected from both static and non-static environment in Figure~\ref{fig:dy1} (bottom),
from which we see the variation of the $S(t)$ accurately captures the dynamics of the wireless channels, because the normalization operation to compute similarity essentially removes the effect of AGC and thus $\varepsilon_{1}(t)$ is removed.
We note that, the similarity $S(t)$ is affected by CSI sampled at two time point, so its value may also vary when the sampling interval varies, adding another unpredictable factor.
In our implementation, we introduce a reference vector $\vec{r} = \left[1,\ldots,1 \right]$
and derive $S(t)$ as the similarity between the $\hat{H}(t_n)$ and the reference vector. We use the variance of S(t) across 0.1s and above a threshold $\Gamma$ to detect the human movement.

\subsubsection{Segmenting Fall-like Activities}
To forge efficient online detection and relatively consistent feature extraction, 
we propose a heuristic algorithm to segment fall-like activities from continuous monitored RF signals.
The key observation is that a fall and fall-like activity (e.g. Sit, Jump and Squat) 
usually comes to a full pause at the end of the motion before transitioning to next movement, 
which may be due to the direction change of the movement (vertical to horizontal).
Similar observation has been reported in previous studies with WiFi~\cite{wang2016rt} and RFID ~\cite{chen2020rf} signals as well.
Therefore, we segment $S(t)$ in a backtracking manner from an observation of motion pause, which is easier to capture than the actual start of an activity.
Meanwhile, as the channel dynamic is caused by the human movements, we derive an approximate acceleration descriptor $a$ to help further filter out daily movements accompanied by a pause with low-intensity (e.g. walk and stop). In addition, we assume the RF signals collected after a fall are also useful and thus a greedy algorithm is used to keep monitoring the $S(t)$ to window the entire fall-like activity.

\begin{figure}[t]
    \begin{subfigure}[b]{0.495\linewidth}
		\includegraphics[width=\linewidth]{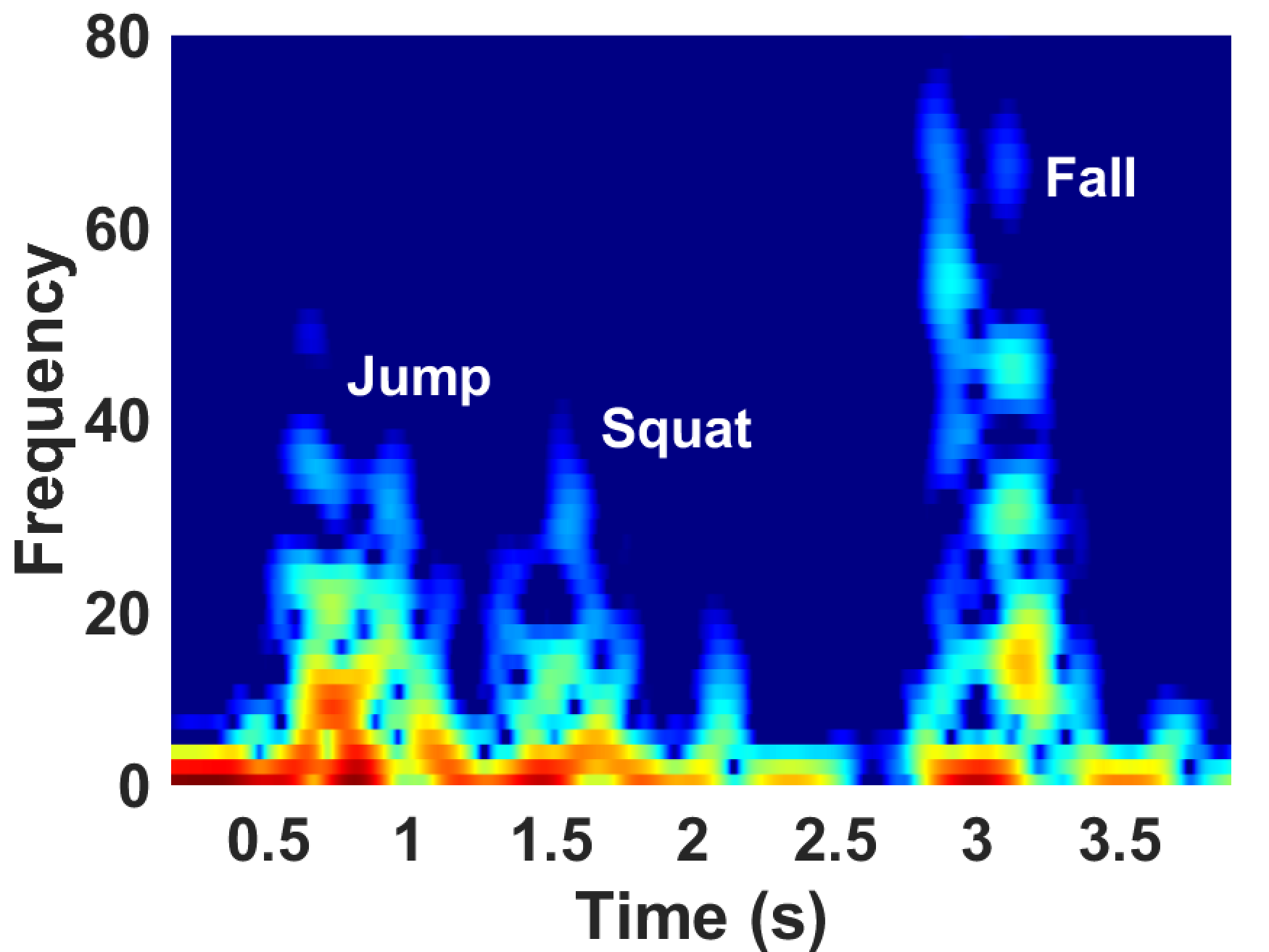}
		\centering
		\caption{Spectrum}
		\label{fig:stftnew}
	\end{subfigure}
	\begin{subfigure}[b]{0.495\linewidth}
		\includegraphics[width=\linewidth]{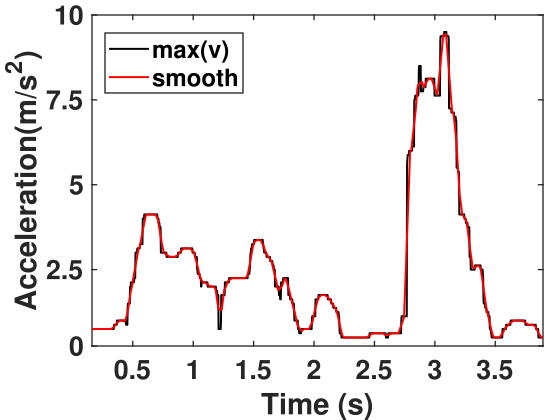}
		\centering
		\caption{$a(t)$}
		\label{fig:PLCR}
	\end{subfigure}
    \caption{The STFT spectrum and the corresponding acceleration of channel dynamic.}
    \vspace{-0.3cm}
    \label{fig:PLCRs}
\end{figure}

The approximate $a$ is computed by using the relationship~\cite{qian2017widar}:
\begin{equation}
a(t)=\frac{\mathrm{d}^{2}}{\mathrm{dt}^{2}} S(t)=\lambda \frac{\mathrm{d}}{\mathrm{dt}} f_{D}(t)
\end{equation}
where $\lambda$ is wave-length of the subcarrier wave, $f_{D}(t)$ is the Doppler frequency shift. We approximate $\frac{\mathrm{d}}{\mathrm{d} t} f_{D}(t)$ by computing STFT of $S(t)$ as STFT is used to capture the frequency component in a small time duration and the frequency component change is caused by the relative movement between transceivers and the reflecting human body. 
Denote the STFT spectrum as $\mathcal{S} \in \mathbb{R}^{F \times T}$, where $F$ is the fix frequency bins and $T$ is the number of time bins.
For each time bins, we have a vector of approximate $\frac{\mathrm{d}}{\mathrm{d} t} f_{D}(t)$ denote as $\vec{v}$, $\vec{v} \in \mathbb{R}^{F}$. We search $max(v)$ as the function of indices of frequency bins that exceed the noise floor via dynamic programming such that:
\begin{equation}
\begin{aligned}
&\operatorname{max(v)}=\underset{f_{1}, \cdots, f_{T}}{\operatorname{argmax}} \sum_{i=1}^{T} \mathcal{S}_{i, f_{i}}, \\
&\text { s.t. }\left|f_{i}-f_{i-1}\right|<=1 ; i=2, \cdots, T .
\end{aligned}
\end{equation}
$max(v) \triangleq\frac{\mathrm{d}}{\mathrm{d} t} f_{D}(t)$ so $a(t)$ may be obtained by calculating $\lambda max(v)$. We consider $a(t)>\Theta=2.5$ indicates a potential fall-like activity as the human normal acceleration in walking is less than $2.5m/s^2$~\cite{yoon2012novel}.
Figure~\ref{fig:stftnew} is the STFT spectrum derived from $S(t)$ contained in  Figure~\ref{fig:dy1} and Figure~\ref{fig:PLCR} is the $max(v)$ derived from the STFT contained in Figure~\ref{fig:stftnew}. 

When applied in real time, once the variance of $S(t)$ is estimated below $\Gamma$, suggesting a pause after a move, SiFall records the time as $t_{end}$ and then searches if there exists $a(t) > 2.5$ in the past five seconds ($t \in [t_{end}-5,t_{end}]$) and records the $max(a)$ and its corresponding time as $t_{max}$.

An online greedy change point detection algorithm~\cite{killick2012optimal} is applied to continuously update $t_{end}$ for one second afterwards to obtain $t_{end}^{*}$: 
\begin{equation}
\mathcal{C}\left(S(t_{end}:t_{end}^{*})\right) +\beta<\mathcal{C}\left(S(t_{end}:t_{end+1}^{*})\right)
\end{equation}
where $\mathcal{C}$ stands for the error of the linear regression and $\beta$ is a penalty value. The rationale behind is that the RF signals collected after the fall-like activity may also contain useful information for identifying the fall.

In the end, SiFall extracts $S(t)$ between $\left [ t_{max}-3s, t_{end}^{*} \right ]$ and performs STFT on $S(t)$ to obtain the fall-like segments. The additional three-second time before $t_{max}$ is used to include as complete fall-like activity as possible because
we would rather contain redundant signal data as compared to missing any possible important data.
Note that the lengths of STFT segments and their corresponding spectrums are variable because the time between $t_{max}$ and $t_{end}^{*}$ depends on the duration of the captured activity. Following that, the STFT segment of the fall-like activity is supplied to the neural network in the back-end for affirmative fall detection. Algorithm~\ref{alg:seg} defines the whole backtracking segmentation process.

\begin{algorithm}

\caption{Fall-like Segmentation Algorithm}
\label{alg:seg}
\SetKwProg{generate}{Function \emph{generate}}{}{end}
\KwIn{$S(t)$; Threshold: $\Theta,\Gamma$; Penalty: $\beta$; $fs$}

\If{movstd(S(t),fs/10) < $\Gamma$;}
{
record $t$ as $t_{end}$, S=STFT([S(t-5fs),...,S(t)])\;

    \If{$max(v)$ > $\Theta$}
    {
    record $t_{max}$\;
    \While{$t$ < $t_{end}$+$fs$} {err($t$)=$\mathcal{C}\left(t_{end}:t \right )$\;
    \If{err(t) > err(t-1) + $\beta$}{$t_{end}^{*}$=$t-1$}
    }
    continue\;
temp = $[S(t_{max}-3fs), S(t_{end}^{*})]$\;
seg = STFT(temp)\;
    }
}

\end{algorithm}

\vspace{-0.3cm}
\subsection{FallNet Design}
\begin{figure*}[h!]
  \centering
  \includegraphics[width=\linewidth,scale=0.75]{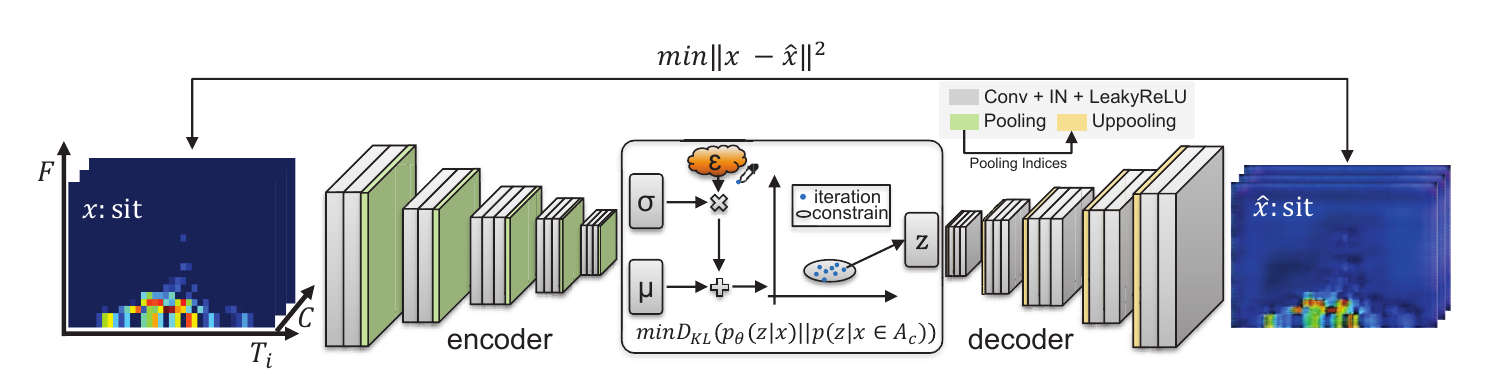}
  \caption{The FallNet Architecture: the encoder, decoder and bottleneck layer.}
  \label{fig:fallnet_architecture}
\end{figure*}

The difficulty now lies in identifying ongoing falls from those RF clips of fall-like activities.
This section elaborates on the design of FallNet, which is able to further identify falls from the RF clips of fall-like activities.
Specifically, we learn the complicated distribution of normal fall-like activities by a variational auto-encoder based FallNet. When used for inference, the FallNet is only able to accurately reconstruct the normal fall-like activities but not real human falls.
We, therefore, identify the activities that result in large reconstruction error as falls.
We first construct the core encoder-decoder architecture of FallNet, which does not rely on data annotation and is able to accept unstructured input data. Then, we elaborate on some special designs of FallNet to cope with particular issues. Finally, we import the variational inference technique to FallNet to make it more generalizable.

\subsubsection{FallNet Architecture}
We design FallNet based on autoencoder architecture which is a well-known deep learning framework to compress data without labels.
The ability to compress data shows its high ability to understand the intrinsic relationship between the compressed data and the original data, hence, a trained encoder is also widely used as a feature extractor.
We train the encoder-decoder only based on the fall-like STFT segments collected from daily activities so the FallNet learns the representative features of daily activities.
When used for inference, the encoder-decoder is able to fully reconstruct the signals of those repeatedly seen normal activities.

\parahead{Encoder}
The input to the network is the signal clip of the $i$th activity $x_i\in \mathbb{R}^{F \times T(i) \times C}$ from a total number of $N$ activities,
where the $F$ is a chosen frequency resolution of the STFT image, $T(i)$ is the time duration of the activity, which might vary across activities, and
$C$ is the number of spatial streams(between Tx and Rx antennas). 
Therefore, the complete information of the three domains, i.e., time, frequency and spatial, are fed into the FallNet.
The encoder of the FallNet learns a nonlinear transformation $\mathcal{F}_\mathcal{E}: \mathit{X} \rightarrow \mathit{Z}$ that maps
the original data space $\mathit{X} \subseteq \mathbb{R}^{m(i)}$ with variable dimensions and 
inconsistency to a compact latent feature space $\mathit{Z} \subseteq \mathbb{R}^{n}$ with uniform dimension. 
$m(i)$ denotes the flattened dimension of $x_{i}$ and $n$ represents the dimension of the latent space of features that are most representative to describe the activities such that:
\begin{equation*}
\textit{z}=\mathcal{F}_\mathcal{E}\left(\textit{x}, \mathit{\Theta}_{\mathcal{E}}\right)
\end{equation*}
where $\mathit{\Theta}_{\mathcal{E}}$ is a set of parameters of the encoder. As the encoder learns the most representative features and automatically filters out the redundancy, it works well with the STFT segments, which may be longer than the actual activities.

The ability of the encoder to project variable-length data space into a uniform latent feature space is owing to our fully convolutional network structure design of the building blocks. The convolution operation itself intrinsically can cope with input of varying lengths, although many people don't notice this because the convolution operation is usually used to process images that are of same length. 
The convolution operator in fact works on local tensor regions and depends only on relative spatial coordinates determinated by the convolution kernel size~\cite{goodfellow2016deep} (refer to Appendix B for more details). As a result, when using the "same padding"~\cite{goodfellow2016deep} in a convolution layer, for an input with dimension $F\times T(i) \times C$, the output will be with the dimension of $F \times T(i) \times C'$, where the only change is the channel dimension $C'$, depending on the number of convolution filters.
In particular, the encoder of FallNet consists of five building blocks of decreased size that are stacked together. Each building block consists of two convolution layers with instance normalization (IN)~\cite{ulyanov2017improved}, an activation function of LeakyReLU~\cite{xu2015empirical}, and a max-pooling layer. All convolution layers fix the convolution filter size to 3 which simulates a larger filter while keeping the benefits of smaller filter sizes in order to reduce the computational overhead \cite{simonyan2014very}. IN is used to cope with the antenna imbalance issue. LeakyReLU is the activation function to bring in non-linearity ability of the network and it can avoid the dying ReLU problem. Max-pooling is used to achieve translation in-variance over small spatial shifts in the input tensor~\cite{szegedy2015going}. The max-pooling layer will decrease the size of the input to half so that the final output size of each building block is $F/2 \times T(i)/2 \times C'$. At the end of the five building blocks, we first average pooling the feature values along the time dimension with an index to record its dimension. Note that $C'$ is determinated by the number of convolution filters which is controlled by us and the $F$ is fixed, a fully connected layer hence can be used to conduct channel-wise linear transformation to map the tensor to a fixed-length vector $z$ with $n$ dimension that represents the extracted features. 

\begin{figure}[t]
\includegraphics[width=0.9\linewidth]{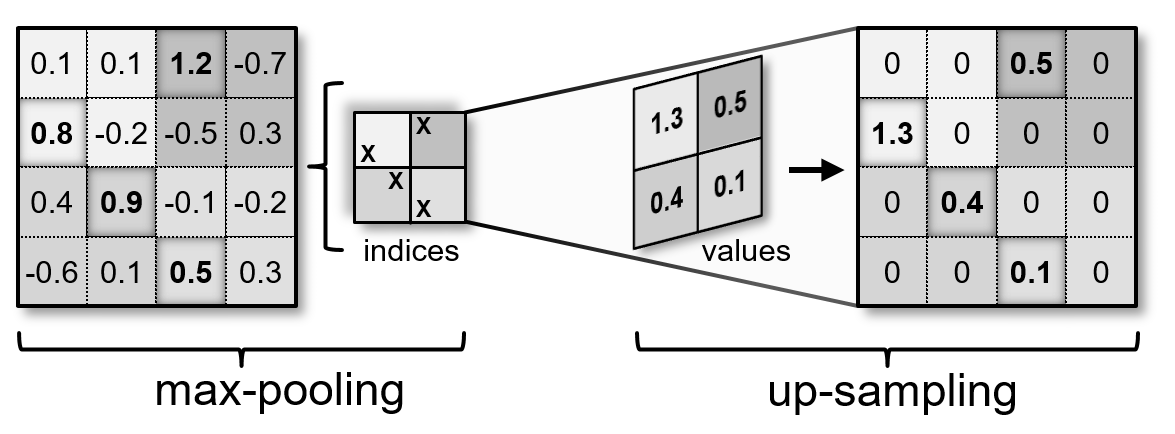}
\vspace{-0.3cm}
\caption{Up-pooling diagram.}
\label{fig:pooling}
\vspace{-0.3cm}
\end{figure}

\parahead{Decoder}
The decoder learns to reconstruct the input signal $x_i$ from the output $z$ of the encoder, such that
\begin{equation*}
\hat{\textit{x}}=\mathcal{F}_\mathcal{D}\left(\textit{z}, \mathit{\Theta}_{\mathcal{D}}\right)
\end{equation*}
where $\hat{\textit{x}}$ is the reconstructed signal, and $\mathit{\Theta}_{\mathcal{D}}$ is a set of parameters of the decoder.
To reconstruct $\hat{x}$, the decoder needs up-sampling oprations to map $z$ back to the size $m(i)$ of the original input smaple $x(i)$. In consequence, the decoder and the encoder are symmetric with the same number of building blocks, except that the max-pooling layers at the encoder is replaced by up-pooling layers at the decoder. As up-pooling~\cite{badrinarayanan2017segnet} utilizes the 2-bit indices stored during max-pooling operation in the encoding phase and up-samples the feature map by filling the values directly to the index position and zero-padding the remaining positions. It avoids parameter learning to reduce the computation overhead. Figure~\ref{fig:pooling} illustrates the up-pooling operations.Another small detail is that the decoder first uses the record index from the previous average pooling operation to zero-pad the $z$ back to the dimension before the fully connected layer, then goes through the five identical building blocks of the decoder.

Consequently, the goal of the FallNet is to learn the parameter sets of encoder and decoder satisfying:
\begin{equation*}
\left(\hat{\mathit{\Theta}}_{\mathcal{E}}, \hat{\mathit{\Theta}}_{\mathcal{D}}\right)=
\underset{\mathit{\Theta}_{\mathcal{E}}, \mathit{\Theta}_{\mathcal{D}}}{\arg \min } \mathbb{E}_{x \sim \mathit{X}}\left[\|\textit{x}-\mathcal{F}_{\mathcal{D}}\left(\mathcal{F}_{\mathcal{E}}\left(\textit{x}, \mathit{\Theta}_{\mathcal{E}}\right), \mathit{\Theta}_{\mathcal{D}}\right)\|^{2}\right]
\end{equation*}

It is worth noting that this learning process only needs the input sample $x$ and does not require any labelled data, therefore, it can benefit from substantial and easily accessible RF samples of daily activities.

\subsubsection{Coping with Antenna Imbalance.}
CSI collcted from different antennas may have different amplitudes, which lead to the imbalance of the power of STFT spectrums. The removal of AGC impact in the CSI denoising phase further amplifies this issue.
As the $C$ channels of input tensor corresponds to different Tx-Rx antenna streams, the FallNet adopts IN that normalizes the antenna streams with learnable affine parameters $\gamma$ , $\beta$ to cope with the antenna imbalance:
\begin{small}
\begin{equation*}
\mathrm{IN}_{\gamma, \beta}\left(X\right) \equiv \gamma \widehat{X}+\beta, \text { where } \widehat{X}=\frac{X-\mu}{\sqrt{\sigma^{2}+\epsilon}}
\end{equation*}
\end{small}
where $\mu$ and $\sigma^{2}$ are computed across spatial dimensions independently for each channel so that every spectrum has the same range of values. $\epsilon$ is a small constant added for numerical stability. Noted that the FallNet removes the commonly adopted Batch Normalization (BN), as the data samples in our case are generated online and may follow different distributions. IN has the same characteristics as BN does, which helps the entire neural network to alleviate gradient saturation and accelerate convergence~\cite{ioffe2015batch} (refer to Appendix A for more details).

\subsubsection{Coping with RF Data Scarcity}
Although the training of the FallNet is free from data annotation, making it possible to  continuously learn from daily fall-like activities, it is not realistic to enumerate all possible fall-like activities. Besides, some types of activities may be relatively dominant owing to specific user activity patterns. As a result, FallNet may be prone to be overfitted. To make the FallNet resistant to such overfitting and be  generalized to function properly, instead of using a vector $z$ with $n$ dimension to represent the learned fall-like activity features, we adopt a bottleneck layer with stochastic sampling operation to make the FallNet become probabilistic.

The reason for doing this is based on our observation (Figure~\ref{fig:stft}) that fall-like activities of the same type are similar, though not identical. By introducing this prior knowledge, we can construct the obtained samples with certain distributions and assume that the same type of activities come from the corresponding distribution to obtain more general sample characteristics. We, therefore, import such prior knowledge into the network, allowing the neural network to learn more generalizable features from the limited data. In particular, we assume that each of the $n$ features of the RF signals follows a normal distribution due to different body shapes or orientations. Refer to Figure~\ref{fig:stftfall} to see that the same actions performed by a single person or multiple persons 
have similarity due to the kinematic consistency. Thus, in the feature space, samples from each normal activity group $A_{c}$ are supposed to follow an $n$-dimensional Gaussian distribution as the activities from the same group (e.g., sit, bow, or jump) are repeated and controlled. We denote a certain activity group as $A_{c}$ with number of $j^{(c)}$ samples. Ideally, all normal samples from different daily activities together form a mixture distribution of Gaussian. 

With such prior knowledge, we therefore impose the constraint to FallNet's learning process and force it to learn a mixture Gaussian distribution over the latent feature space, rather than learning a vector of feature representations $z$ that may be over-fitted with limited data samples. To this end, we modify the output of the encoder from $z$ to two vectors $\mu_{c}$ and $\sigma_{c}$ that represents mean and variance of the activity group Ac that each training sample belongs to, respectively, where $\mu_{c}, \sigma_{c} \in \mathbb{R}^{n}$, $n$ is the number of features. The FallNet learns the two vectors to parameterize the feature distribution of $A_{c}$. A constrain loss is added to minimize the Kullback-Leibler (KL) divergence between the learned disrtibution of the parametric representation and the desired distribution  $p\left ( z | x \in A_{c} \right )  \sim \mathcal{N}\left(\mu_{c}, \sigma^2_{c}\right)$ such that: 
\[\mathit{L_{c}} = -\frac{1}{2}\sum_{i=1}^{n}\left ( \mu^{2}{\left ( i \right )} + \sigma^{2}{\left ( i \right )} - \log \sigma^{2}{\left ( i \right )} - 1 \right )\]
where $\mu(i)$ and $\sigma(i)$ denote the $i$-th element of the $n$-dimensional vectors $\mu$ and $\sigma$. In such a way, each activity is modeled as a multivariate Gaussian distribution with $n$-dimensional features in the latent space. Different activities have different mean vectors $\mu_{c}$ and variance vectors $\sigma_{c}$ to represent different Gaussian distributions. As the number of samples increases, the hidden space gradually forms a complex Gaussian mixture distribution: 
\[p_\theta(z)=\sum^c \frac{j^c}{N} p\left(x \in A_c \mid \mu_c, \sigma_c^2\right)\]
If the latent distribution is valid, correspondingly, any of the latent space samples from the distribution should be able to reconstruct $x$ well. Therefore, the input of the decoder now becomes a $z$ that is stochastically sampled from the corresponding $\mu$ and $\sigma$ such that

\begin{equation*}
\hat{\textit{x}}=\mathcal{F}_\mathcal{D}\left(\delta \textit{z} \sim  \mathcal{N}\left(\mu, \sigma^2 \right), \mathit{\Theta}_{\mathcal{D}}\right)
\end{equation*}

where $\delta$ represents a random sampling operation.
On the other hand, the back-propagation of training neural network requires deterministic operations at each neural network nodes which iteratively pass the gradients and apply the chain rule. The stochastic sampling operation however is not a continuous function and thus not differentiable to obtain the gradient. To make the neural network trainable, the FallNet adopts the reparameterization technique~\cite{kingma2013auto}. It generates random $\varepsilon$ from a standard normal distribution $ \mathcal{N}\left ( 0,1 \right )$ independent of the neural network nodes. The latent sample $z$ is obtained through scaling and transformation by $z = \mu + \sigma \times \varepsilon$. The reparameterization allows $z$ to be sampled from the corresponding distribution of $\mu$ and $\sigma$ at each iteration while the random sampling itself is not involved in the training process. As the sampled $z$ is deterministic at each iteration its gradient can be back-propagated to train the entire neural. Consequently, the objective of the FallNet is revised:
\begin{equation*}
\underset{\mu, \sigma, \mathit{\Theta}_{\mathcal{D}}}{\arg \min } \mathbb{E}_{x \sim  \mathit{X} }\left[\|x-\mathcal{F}_{\mathcal{D}}((\mu_{x}+ \sigma_{x} \times \varepsilon ,\mathit{\Theta}_{\mathcal{D}} )\|^{2}\right], \varepsilon \sim \mathcal{N}(0,1)
\end{equation*}

In addition, the FallNet design also employs data augmentation scheme to compensate the data scarcity and improve model generality. The FallNet imposes two specific augmentation schemes: (i) To simulate a low SNR scenario, before being converted to STFT spectrums, for each segmented $S(t)$, we add Gaussian white noises, which equals to adding noises in the channel domain of the input tensor;
(ii) To alleviate the limitation of time resolution due to the fixed STFT window length. Each input tensor $x_{i}$ goes through three rounds of random horizontal shift~\cite{torch}, with the shifting length smaller than the STFT window length. At the end we are able to fabricate $24\times$ the amount of original data to augment the training size. 

\subsection{Online Detection and Model Updating}

After pre-training with a normal activity dataset $X$, the FallNet has established the distribution of the anchor daily activities in the latent feature space. Let each activity segment $x$ go through the FallNet, we can derive the statistics of reconstruction error of the dataset including its average $\alpha$ and median $\gamma$. In the online detection phase, the FallNet takes the real time segmented STFT samples for inference in a single run, and measures its reconstruction error $e$. If $e > 2\alpha$, it is detected as a fall and at the same time $\alpha$ and $\gamma$ remain unchanged. If $\alpha < e < 2\alpha$, the system takes it as a suspicious daily activity and saves the segmented samples for feature reference, but  $\alpha$ and $\gamma$ are recalculated and updated accordingly. Once the change of $\gamma$ exceeds a threshold, the system takes it as an indication of significant change in the environment. If $e < \alpha$,  the system updates the $\alpha$ and $\gamma$ and then performs data augmentation where a mini-batch of augmented data samples are fed to the FallNet for retraining the model. Therefore, the system keeps evolving with the feedback of reconstruction error $e$ and adaptively updates the threshold $\alpha$ to determine falls.

As the system runs in real-time, the incoming fall-like samples for inference may bring two types of distribution shift, one being the \textit{semantic shift} caused by the individualized movement patterns across people, the other being the \textit{covariance shift} due to environment variation over time. As SiFall eliminates the environment impact by extracting the dynamics of RF signals, the covariance shift is well accommodated along with the continuous update of the FallNet. The saved suspicious daily data samples are utilized to deal with the semantic shift. Whenever an adequate amount of suspicious daily data samples (i.e., 50 as set in our current implementation) are collected, SiFall performs principal components analysis (PCA) to reduce the dimension to $n$ and then performs mean-shift clustering~\cite{pedregosa2011scikit} to identify $N$ clusters. Two criteria are applied to handle the cluster points, namely, representativeness and diversity. We examine the largest cluster as it indicates many repeatable activities which are unlikely to be human falls. SiFall retrieves the signal segment of the centroid of the largest cluster, produces 24$\times$ augmented data, and feeds that to the FallNet for model retraining. SiFall also notices when there is a cluster that is far away from other clusters. The cluster is taken as a potential undiscovered user activity group and its signal segments are kept for later examination when adequate amount of such suspicious data are collected. The remaining signal segments are discarded and the counter is updated till next time the number of saved samples reaches 50.

Based on the above described mechanism of automatic model update, SiFall does not require explicit human intervention for most of the time. Only when a "fall" is detected SiFall triggers an alarm for possible human intervention. The corresponding data samples are saved with a timestamp regardless whether the detected "fall" is a true positive or false positive. The human user may examine the saved "fall" samples at any later time to decide whether they are true positives in which case the samples are discarded, or false positives in which case the samples are augmented and fed back to the FallNet for retraining.

\section{Evaluation}
In this section, we evaluate the performance of \systemname{}. 
We first introduce our experimental settings and then present the results.

\begin{figure*}[h!]
   \centering
   \includegraphics[width=0.95\linewidth]{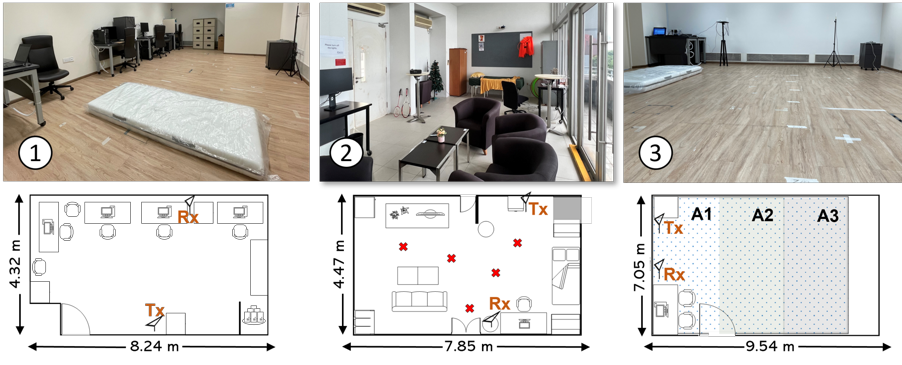}
  \caption{The environment of the three testbeds and their floor plans. }
  \label{fig:testbed}
\end{figure*}
\subsection{Experimental Setting}

We implement SiFall with two commercial off-the-shelf (COTS) APs as the Tx and Rx to collect the WiFi CSI,
one laptop connected to the Wi-Fi receiver to serve as the front-end edge server and one back-end server.
We use a camera to capture the ground truth. 

\parahead{Hardware}
We use COTS COMPEX WPJ558 equipped with Atheros SoC QCA9558 in the experiment.
We let these two APs transmits 200 packets per second on a 20MHz channel in 2.4GHz frequency band.
We fix the Modulation and Coding Scheme (MCS) to reduce packet loss and noises. We use Atheros-CSI-Tool~\cite{xie2018precise} to collect raw CSI data.
The receiver forwards the collected CSI to the ThinkPad T430 laptop with an Intel Core i5-3360M CPU to process the RF signals and generate STFT segments (as introduced in Section 3.1).
We use a Linux desktop computer equipped with Intel Core i9-9820X CPU and one Nvidia 2080Ti GPUs to work as the back-end server to maintain the FallNet and perform real-time inference to detect the falls.

\parahead{Testbed}
We test SiFall based on three testbeds - an emulated "bedroom" with an enclosed space measured 4.32m $\times$ 8.24m for comprehensive evaluation (testbed 1), a real apartment room measured 7.85m $\times$ 4.47m for system adoption test on a daily basis (testbed 2), as well as a big open area measured 9.54m $\times$ 7.05m to test the effective sensing range of the system (testbed 3).
Figure~\ref{fig:testbed} depicts the three different testbeds. The marked Tx and Rx indicate the locations of the WiFi Tx and Rx antennas.

\parahead{Ground Truth}
We use a camera to record the detailed human activities at a frame rate of 30fps, and manually analyze the recorded video clips to generate the ground truth.
We use network time protocol (NTP) to synchronise the time in the camera recordings and the collected Wi-Fi CSI data.

\parahead{FallNet Pretraining}
We pre-train the FallNet with the data collected intermittently during 3 months in testbed 1, including 1447 sets of
STFT segments of sitting, jumping, swinging, bowing, running, and other daily activities, 
augmented 24 times to produce a total number of 34,728 samples. 
Correlation among raw samples is removed by OpenCV, and the weight parameters are initiated by kaiming initialization~\cite{he2015delving}. 
The model was trained by Adam~\cite{kingma2014adam} optimizer on 4 Nvidia 2080Ti GPU for 2 hours.

\parahead{Testing Subjects}
We recruit 16 volunteers (11 males and five females) 
with ages between 21 and 56 to take part in our experimental evaluation (with IRB approval).
Table~\ref{t:tester} summarizes the detailed information of all volunteers. The testing subjects are highly diverse in their age, weight, and height.
Specifically, the body weight of our volunteers varies from 42kg to 100kg. 
Their body height varies from 155cm to 186cm, and their age varies from 21 to 56 years old. 

\begin{table}[h!]
\centering
\footnotesize
\addtolength{\tabcolsep}{-4pt}
\caption{Summary of the testing subjects}
\label{t:tester}
\begin{tabular}{lcccccccccccccccc} 
\hline\hline
\textbf{\textit{\#Person}}          & \textit{1}   & \textit{2}  & \textit{3}  & \textit{4}  & \textit{5}  & \textit{6}  & \textit{7}  & \textit{8}   &  \textit{9}   & \textit{10} & \textit{11} & \textit{12} & \textit{13} & \textit{14} & \textit{15} & \textit{16}  \\ 
\hline\hline
\textbf{Age}                        & 34           & 25          & 21          & 25          & 28          & 27          & 26          & 25      
         & 29           & 27          & 25          & 29          & 26          & 22          & \textbf{52} & \textbf{56}   \\
\textbf{Height(cm)}                 & 165          & 167         & 177         & 188         & 184         & 171         & 173         & \textbf{155} &  \textbf{186} & 173         & 165         & 175         & 172         & 166         & 173         & 155    \\
\textbf{Weight(kg)}                 & 62           & 52          & 65          & 85          & 73          & 61          & 74          & \textbf{42}  & \textbf{100} & 65         & 52          & 71          & 63          &53         & 60          & 62   \\
\textbf{Gender}                     & M            & F           & M           & M           & M           & M           & M           & F   & M            &    M        & F           & M           & M           & F           & M           & F      
\end{tabular}
\end{table}

\parahead{RT-Fall}
We compare the performance of \systemname{} with RT-Fall~\cite{wang2016rt}, 
which is, to the best of our knowledge, the only RF-based fall detection system which claims being able to achieve real time fall detection in practice. 
RT-Fall identifies fall-like activities based on a pre-defined threshold on the measured CSI phase difference between two Rx antennas and segments the collected CSI stream with a fixed 3s time window.
RT-Fall then feeds the derived statistical phase and amplitude features of the CSI segment into a pre-trained SVM model to identify falls. 
We reproduce the system and train an SVM classification model of RT-Fall with the data collected from our testbed, the same as what we use to pretrain our FallNet. 

\subsection{End-to-end Evaluation}
We first conduct intensive movement experiments with 12 subjects and report the end-to-end performance. After that, the proposed system components are evaluated based on the detailed experiment results.

\subsubsection{Methodology}

12 testing subjects (\#P1,\#P6-\#P16) are involved to conduct the experiment in a sequential order. 
Each testing subject is requested to move freely around one and half an hours inside the bedroom testbed as depicted in Figure~\ref{fig:testbed}.
We request each of them to perform the following actions at their will when they move around:
"jump", "squat", "sit to the floor", "sit to the chair", "knee down", and "bow" at least three times at different locations and with different body orientations. Other than the requested type of movements, they are free to perform any other activities at their will. We summarize other fall-like movements that are hard to quantify as "swing".

\begin{table}
\centering
\footnotesize
\addtolength{\tabcolsep}{-4pt}
\caption{Types of Falls}
\label{t:falls}
\begin{tabular}{cll} 
\hline\hline
~\textbf{Types~~} & ~ ~ ~ & \textbf{Examples} \\ 
\hline\hline
~ "walk fall"~ ~ & ~ ~ & \begin{tabular}[c]{@{}l@{}}slip, stumble\\scenes: rushing to answer the telephone, slipping in the bathroom, \\and tripping over the cable, etc.\end{tabular} \\ 
\hline
~ "stop fall"~ ~ & ~ ~~ & \begin{tabular}[c]{@{}l@{}}lost balance, lost consciousness\\scenes: coming out of bed, epileptic seizure, stroke,\\and heart attack, etc.\end{tabular} \\ 
\hline
~ "slow fall"~ ~ & ~ ~ ~ & \begin{tabular}[c]{@{}l@{}}dizziness/vertigo, weakness\\scenes:~arthritis pain, transfer to a dim room, postural hypotension, \\and vision disorder, etc.~ ~~\end{tabular} \\
\hline
\end{tabular}
\end{table}
To mimic unconscious falls as much as possible while meeting the IRB requirement on risk control, we set up a safety mattress and experiment with the falls of three categories~\cite{attar2021common,rubenstein2006falls,kwan2014assessment}: (1) for "walk fall", the subject is asked to walk around the mat and instantly fall on the mat once a random alarm is triggered by us - the fall is performed regardless the instant body orientation of the testing subject; (2) for "stop fall", the subject stands still on the mat and tries to dodge the tennis balls thrown at her - if she happens to fall the activity is noted as a valid "stop fall", and as swing activity otherwise; (3) for "slow fall", the subject keeps standing still until we give a random alarm when she simulates a slow fall on the mat. Table~\ref{t:falls} illustrates the three categories of falls with corresponding real life scenes and examples. It is worth noting that regardless of the type of falls, the falling orientation is random during the experiments based on the reaction of the subject. During the experiment, SiFall continuously operates and each of the 12 testing subjects enters the bedroom in sequence. The total experiment duration for all 12 testing subjects is about 19.3 hours. The FallNet model is continuously maintained and updated throughout the experiment. We evaluate the performance with True Positive Rate (TPR) and False Positive Rate (FPR) metrics, where TPR is true falls out of SiFall reported falls and FPR is falsely reported falls out of other activities. The accuracy is calculated by the percentage of correctly detected falls and non-falls against the ground truth.

\subsubsection{Overall Performance}

\begin{figure}[t!]
\includegraphics[width=\linewidth]{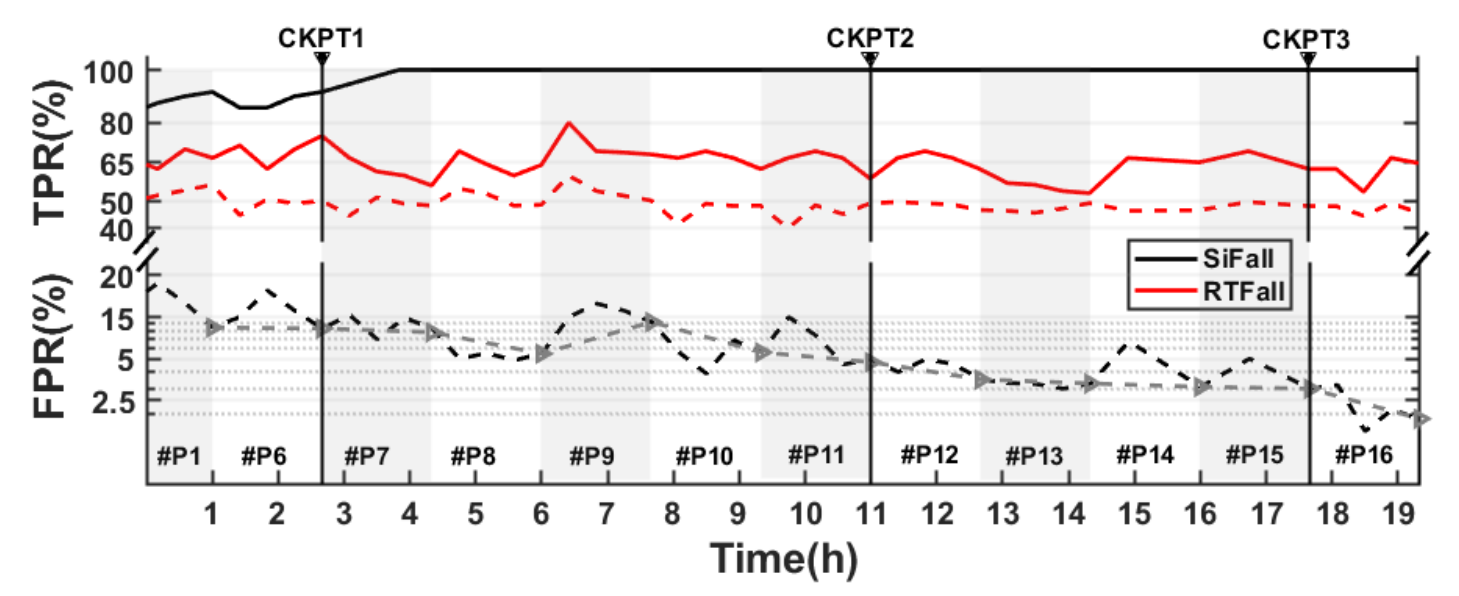}
\centering
    \caption{System end-to-end performance evolution over time across different test subjects.}
    \vspace{-0.3cm}
    \label{fig:tag}
\end{figure}

During the 19.3 hours experiment, SiFall captures a total number of 1497 fall-like activities, of which 523 segments are intentional activities performed by the testing subjects (including 123 falls and 400 required fall-like activities). Among the 123 falls, 60 are "walk fall", 33 are "slow fall" and 30 are "stop fall". We derive the TPR and FPR in about every 20 minutes and plot the results over time in Figure~\ref{fig:tag}, where TPR is represented by the black solid line and FPR is represented by the balck dashed line. 
Both the TPR and FPR vary over time as the FallNet model continuously evolves when more training data are collected from the testing subjects. We see a clear trend of improvement on both the TPR and FPR.

First, the TPR improves quickly over time. 
From ~83\% at the beginning of the experiment, the TPR constantly improves over time and reaches 100\% within 4 hours of operation, which demonstrates SiFall's capability in accurately identifying the abnormal falls from normal daily activities.
Second, the FPR of \systemname{} improves greatly over time.
The falsely reported falls by \systemname{} are ~6.7 per hour in the first two hours and eventually drops to below 1 per hour in the last two hours of the experiment.
While the TPR shows a clear trend of improvement over time, the FPR occasionally fluctuates, especially during the experiment of each individual testing subject. That is mainly due to the fact that our experiment does not restrict how each testing subject performs certain activities, and as a result some testing subjects may choose to perform more activities similar to falls, and in different orders. For example, one (\#P9) prefers challenging SiFall system by performing more "sit on the floor" activity which is more similar to "slow falls" and results fluctuated FPR during his experiment. If we focus on the FPR statistics by the end of each testing subject's experiment (the gray line), we may see steadily improved performance. At the end of the 19.3 hour experiment, SiFall is able to achieve 100\% TPR and 1.8\% FPR.

We simultaneously run RT-Fall for comparison and plot the achieved TPR and FPR of RT-Fall in red in Figure~\ref{fig:tag}. We find that during the real time operation RT-Fall achieves a much lower performance, with its TPR of 64.9\% and FPR of 49.2\%. Since RT-Fall does not have the ability to self-evolve, it cannot gain performance over time and it fluctuates across different testing subjects. Overall the comparative results show huge comparative advantage of SiFall over the SOTA available real-time RF fall detection approach.

\begin{figure*}[t]
	\begin{subfigure}[b]{0.495\textwidth}
		\includegraphics[width=\textwidth,height=5.5cm]{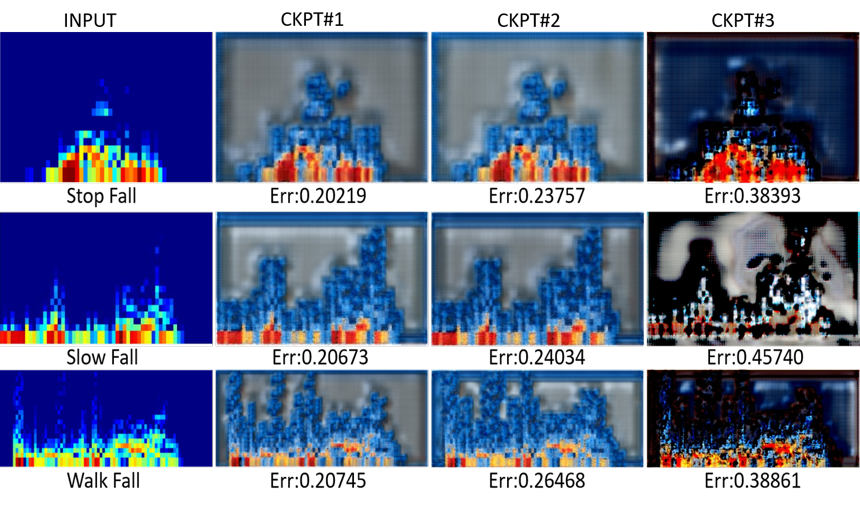}
		\centering
		\caption{The original and reconstructed STFT segments of Falls. }
		\label{fig:netfall}
	\end{subfigure}
    \begin{subfigure}[b]{0.495\textwidth}
		\includegraphics[width=\textwidth,height=5.5cm]{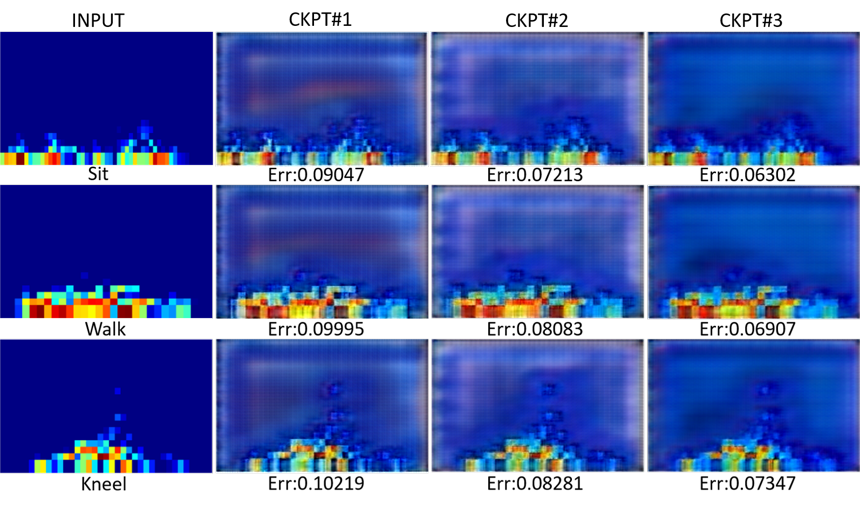}
		\centering
		\caption{The original and reconstructed STFT segments of other activities.}
		\label{fig:netnormal}
	\end{subfigure}
    \caption{Visualization of the FallNet original input and reconstructed output STFT segments.}
    \label{fig:fallnet}
\end{figure*}

\subsubsection{FallNet Visualization}
We visualize the FallNet input and output to demonstrate the rationale when applying FallNet to detect the falls.
Specifically, we impose three checkpoints during the experiment (as indicated in Figure 9 as CKPT1 to CKPT3, after the test of subject \#P6, \#P11, and \#P15, respectively). At each checkpoint, we freeze the FallNet model and memorize it for detailed investigation. We feed different RF signal segments collected from normal activities and falls into the restored FallNet models at the three checkpoints, respectively, to examine the reconstructed output from the FallNet.
In Figure~\ref{fig:fallnet}, we plot the STFT segments of the input signal and the reconstructed STFT segments by the FallNet for different types of falls (Figure~\ref{fig:netfall}) and other ordinary activities (Figure~\ref{fig:netnormal}). We clearly see that while most STFT segments of most ordinary activities can be recovered by the FallNet the STFT of falls cannot. We also observe that as the model evolves (from CKPT1 to CKPT3), the reconstruction errors of all the falls increase while the reconstruction errors of ordinary activities decrease. Note that the errors is computed as the L2-norm of the difference between the FallNet input and output. The reconstruction errors of falls are order of magnitude higher than those of ordinary activities.

The visualization suggests that the FallNet is able to continuously learn better latent distribution to describe the human daily activities and based on that make more accurate detection of falls as outliers. Notably the low-frequency part of the spectrum and the end-of-motion part remain clear despite the deteriorating quality of the reconstructed STFT samples of falls, which suggests that the FallNet indeed utilizes the low-frequency and end-of-motion features in discriminating the samples.

\subsubsection{SiFall Segmentation Performance.}
The segmentation algorithm of SiFall depends on detecting the status of human movement and is threshold based. 
We separately evaluate the accuracy of the movement detector and the threshold sensitivity.

\begin{figure}[h]
    \begin{subfigure}[b]{0.495\linewidth}
		\includegraphics[width=\linewidth]{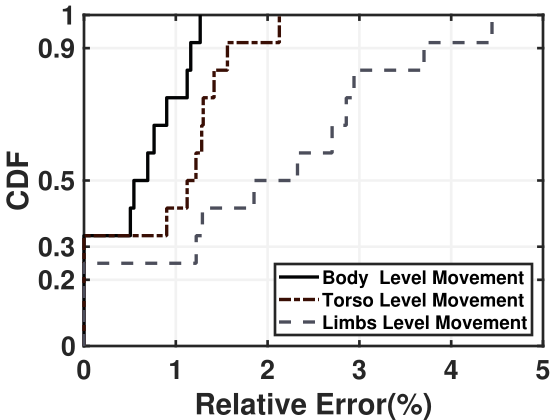}
		\centering
		\caption{}
		\label{fig:move}
	\end{subfigure}
	\begin{subfigure}[b]{0.495\linewidth}
		\includegraphics[width=\linewidth]{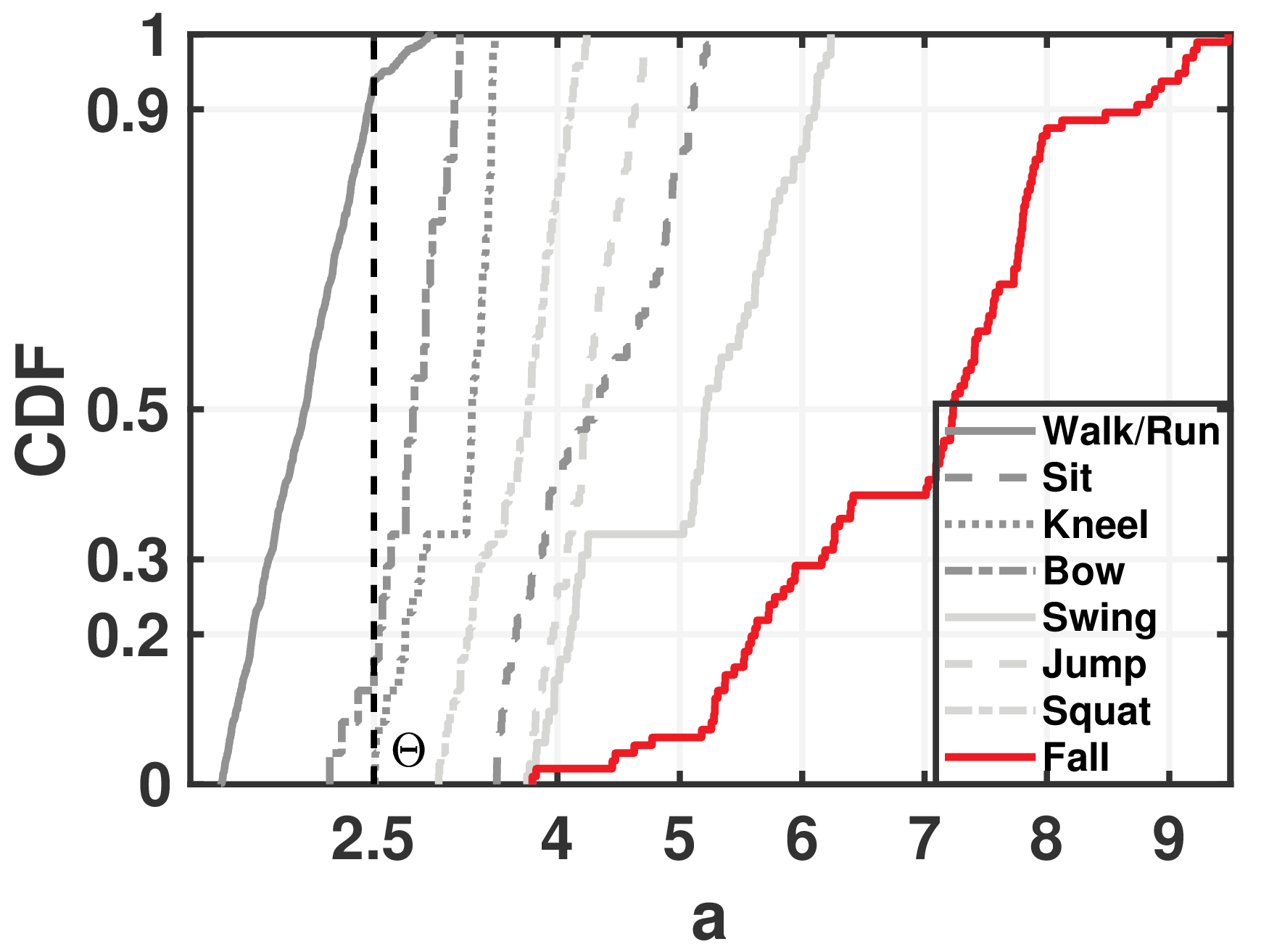}
		\centering
		\caption{}
		\label{fig:velocity}
	\end{subfigure}
    \caption{SiFall (a) segmentation performance of movement detection error and (b) the acceleration distribution of different types of movements.}
    \label{fig:overall}
    
\end{figure}

\parahead{Movement Detection}
We classify the human movement into three levels. 
Body level movement refers to the whole body movement involving position change, 
such as walking and running. 
Torso level movement refers to torso movement without position change, such as bow and squat. 
Limbs level movement refers to limbs and hands movements at minor scale such as shaking hands and typing. 
Figure~\ref{fig:move} reports the percentage of relative error (false negative rate) of the corresponding movement detection results when testing subjects move freely in testbed environment 1. The figure plots the cumulative distribution based on the movements from 12 testing subjects (\#P1, and \#P6-16).
The experiment logs SiFall movement detection performance at body level, torso level and limbs level movement with median error of 0.5\%, 1.1\%, and 1.8\%, and 90th-percentile error of 1.2\%, 1.6\% and 3.7\%, respectively.

\parahead{Threshold Sensitivity}
To quantitatively evaluate the reliability of the threshold $\Theta=2.5$ as fixed in the SiFall implementation, we derive the maximal frequency of each human movement and project to its acceleration. Figure~\ref{fig:velocity} depicts the cumulative distribution of the projected acceleration for different types of movements.

We see that all fall activities have their derived acceleration above the threshold and the majority of other fall-like activities are also captured with the current threshold setting. On the other hand, most ordinary walk and run movements are screened out by the threshold. 12.5\% of bow activities are screened out as well. The result suggests that the threshold setting is effective in screening falls and fall-like activities. It is also obvious that SiFall is robust to the threshold setting - its accuracy will not be impaired when the threshold falls in the range between 2 to 3.7.

\begin{figure}[h]
  \centering
  \includegraphics[width=0.9\linewidth]{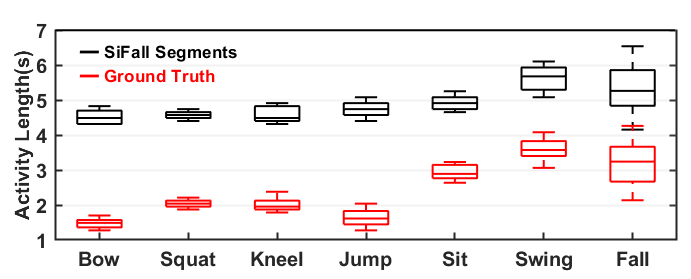}
  \caption{Comparison of the ground truth and SiFall segmented lengths of different activities.}
  \label{fig:seglen}
  \vspace{-0.3cm}
\end{figure}

\parahead{Segmentation Length}
Following the strategy of "more is better than less", the SiFall segmentation algorithm aims at capturing the activities with redundancy in their time durations. Figure~\ref{fig:seglen} compares the lengths of SiFall captured segments and the corresponding ground truth time durations of the activities. 
In general the SiFall segments have an average duration of 5.1s, which is longer than the actual activity duration averaged at 2.6s. Additional 2.5s signal data are included in the SiFall segments for redundancy. Overall, SiFall segmentation algorithm introduces necessary redundancy in extracting the signal segments while maintaining the signal processing overhead on the extra signal durations acceptable.

\subsection{Daily Life Adoption Test}
\begin{figure}[t!]
	\centering
	\includegraphics[width=\linewidth]{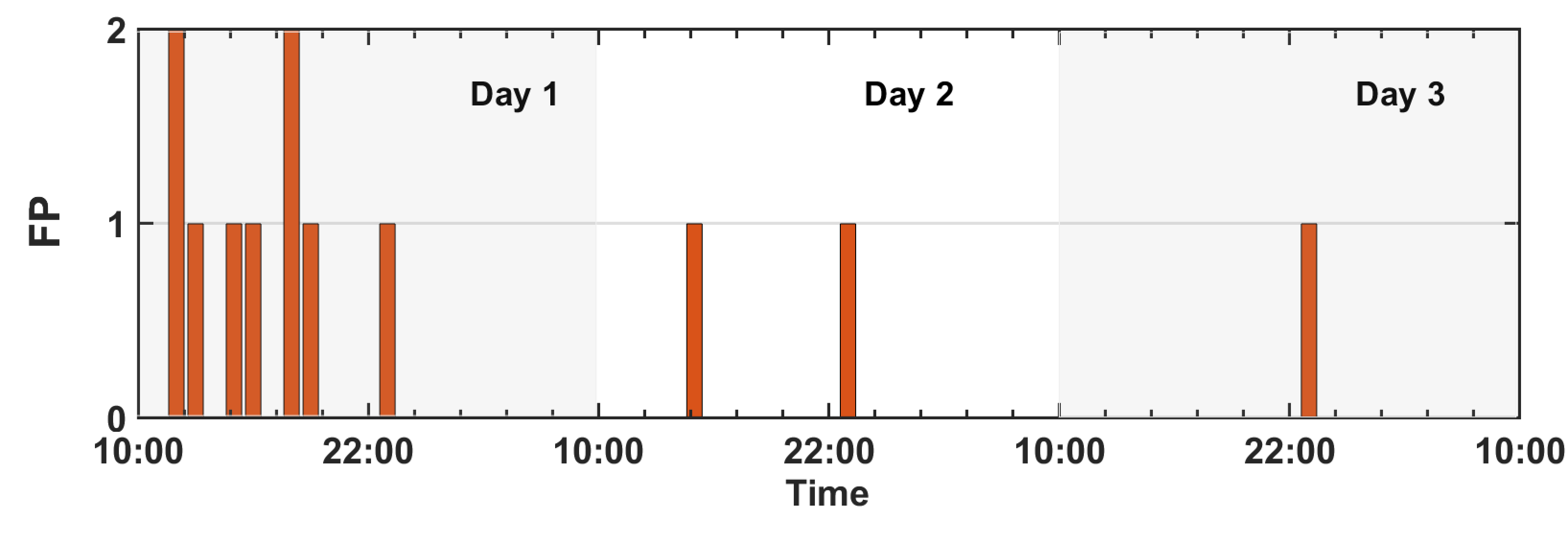}
   \vspace{-0.6cm}
    \caption{False alarms that occur during the daily life adoption test across three days. }
    \label{fig:normal}
    \vspace{-0.3cm}
\end{figure}
In the above experiment, human subjects were asked to continuously perform a large number of falls and fall-like activities in a short time duration for comprehensive evaluation. To further challenge our system and understand the long-term performance in a more realistic setting, we adopt SiFall with pretrained FallNet from the emulated bedroom (testbed 1) to a real apartment room (testbed 2). We conduct a continuous three day evaluation with one testing subject (\#P2) working and living inside the testing room day and night. A total number of 204 fall-like activities (67 on day 1, 63 on day 2, and 74 on day 3, respectively) are captured at the front end and sent to the FallNet for fall inference. Totally 12 false alarms are triggered (9 on day 1, 2 on day 2, and 1 on day3, respectively). Figure~\ref{fig:normal} depicts those false alarms and their occurrence time.
After verifying with the testing subject, the first-day false alarms mainly come from sitting on the sofa and they are phased out gradually with the model update. The first false alarm on the second day is raised when an object falls from the wardrobe and the testing subject picks it up immediately. The second false alarm on the same day is raised when the subject jumps and dives into the sofa from the back of the sofa. The last false alarm on the third day is triggered when the testing subject does handstand on the Yoga mat. We see a clear trend that the false alarms dramatically reduce when the FallNet model is continuously updated over time.
Quantitatively, the false alarm rate decreases from 13.4\% on the first day to 1.4\% on the last day. The experiment results suggest that SiFall has the ability to learn from personalized daily activities, and build evolved models for more accurate fall detection. However, some rarely-seen combination of movements may still trigger the false alarm, which we expect would reduce when SiFall continues to see more repeated occurrences of such activities over longer time of deployment.

After the three day continuous monitoring of the daily activities with SiFall, we perform a purposed experiment to evaluate its detection accuracy of true falls. We freeze the model update of SiFall, and let the testing subject perform ten emulated "stop falls" and "slow falls" following the same methodology illustrated in $\S$4.1. The falls are performed across 5 different locations in the room (shown as "X" in Figure~\ref{fig:testbed}). We then pour a lot of powder on the floor, let the testing subject wear safety gear, keep jogging in the room till five "walk falls" are collected. Apart from those, the testing subject also simulates a fall that rolls from the bed as well as a slipping fall when trying to sit on the office chair. SiFall can detect all the above falls except for the rolling fall from the bed with 94.1\% detection rate, demonstrating the high reliability of SiFall in real-life application.

\begin{figure}[t!]
  \centering
  \includegraphics[width=\linewidth]{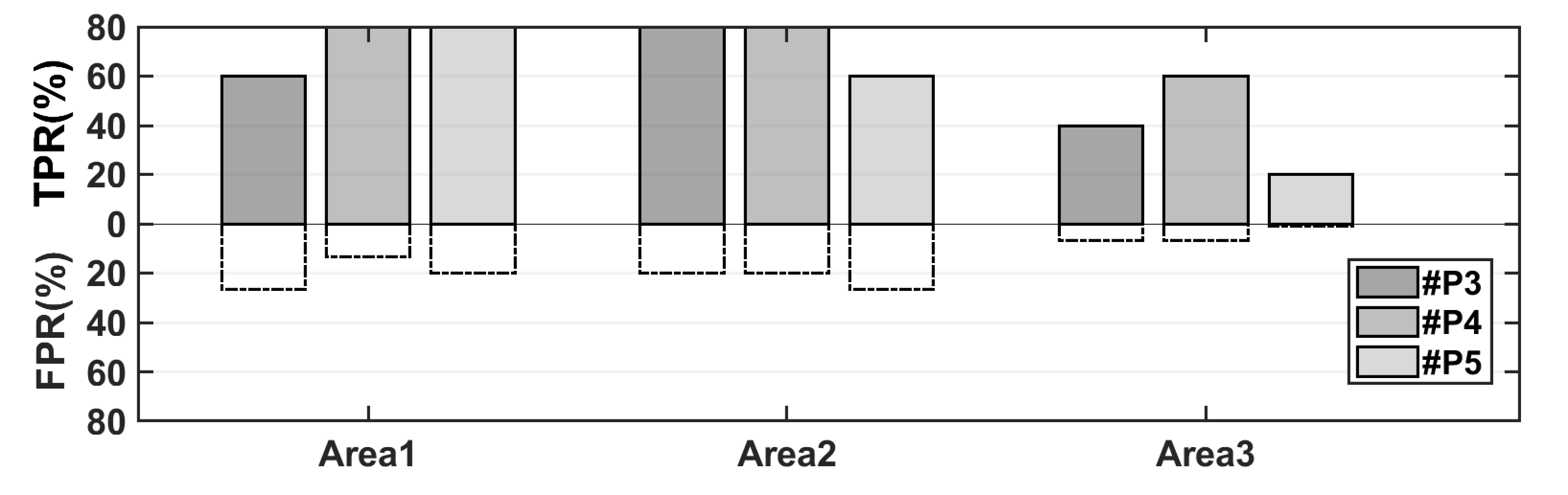}
  \caption{Accuracy of different person at different link distances}
 \vspace{-0.3cm}
  \label{fig:acc3}
\end{figure}

\subsection{Effective Covering Range}
As SiFall captures fall-like activities based on sensing the wireless channel dynamics, we want to evaluate its effective sensing range. We deploy SiFall with the pretrained FallNet from the "bedroom" environment (testbed 1) to a bigger open area (testbed 3) without fine tuning the model.
Three testing subjects (\#P3,\#P4 and \#P5) are requested to perform "jump", "sit to the floor", "swing", and "walk fall" (each for five times with a random sequence) repeatedly in three different areas (as depicted in Figure~\ref{fig:testbed}) with a distance of 1-3 meters, 3-5 meters, and 5-7 meters, respectively, to the LOS link of the Wi-Fi transceivers.
Figure 14 reports the TPR and FPR of the three testing subjects when experimented in the three different areas. We see that SiFall performance is robust when adopted across the environment. The accuracy is reasonably good when the testing subjects move to as far as 5m away from the Wi-Fi link. The average TPR and FPR in area A1 and area A2 were 73.3\% and 20\%, 73.3\% and 22.2\%, respectively. When the distance increases to 7m in area A3, the average TPR drops significantly to 40\% and the FPR drops to 4.4\% at the same time due to failures in detecting and segmenting all fall-like activities. Note that when directly migrating the FallNet model to a new environment (from the "bedroom" in testbed 1 to the big open area in testbed 3), SiFall still achieves an average accuracy of 78.3\% which is impressive. We expect the accuracy will further improve with time when more human activities are captured and consumed by the FallNet model to evolve.

\subsection{Computation Overhead}
\begin{figure}[h]
\centering
\label{f:overall}
    \begin{subfigure}[b]{0.58\linewidth}
    \includegraphics[width=\linewidth]{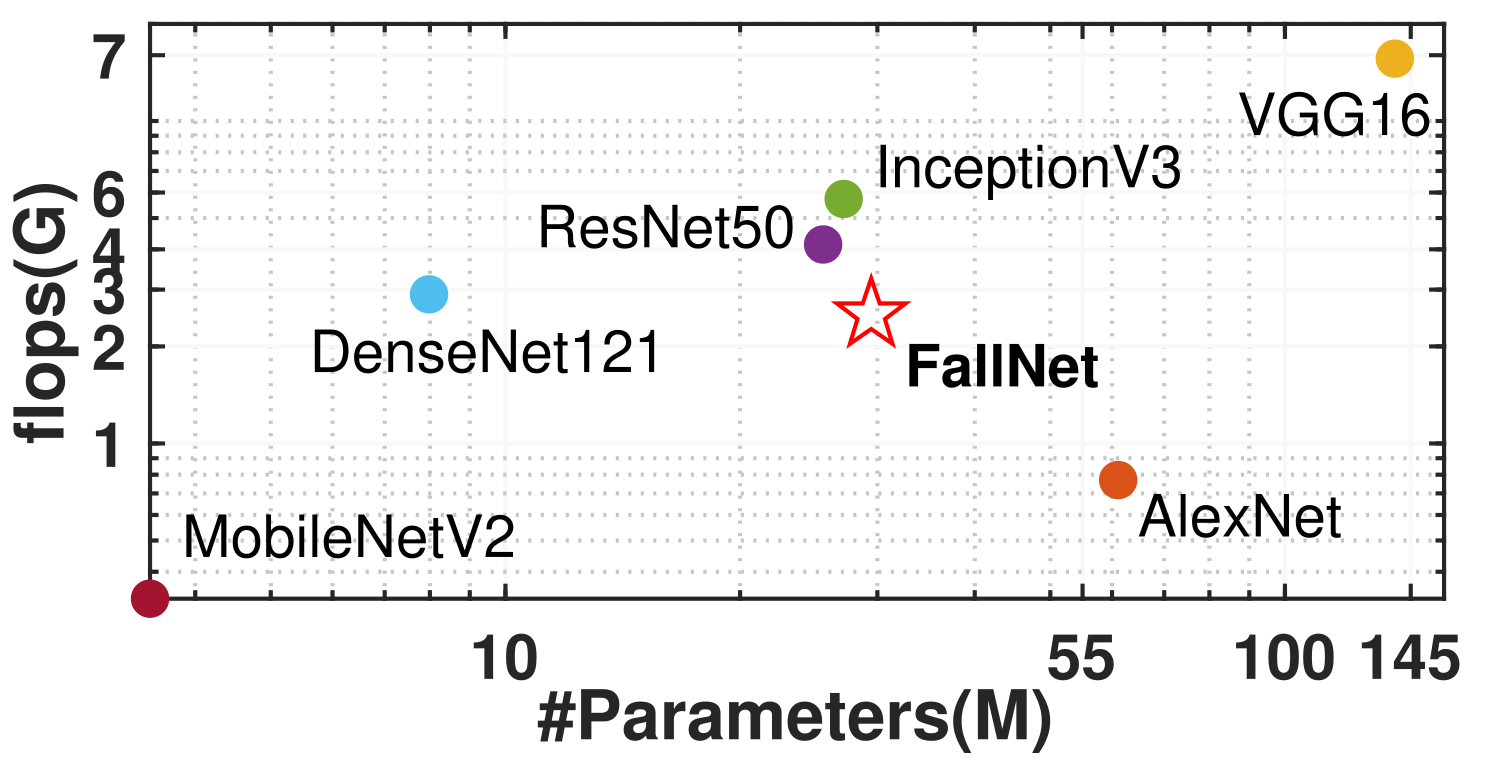}
    \caption{\footnotesize{FallNet model complexity compared with SOTA CNN-based models.}}
    \label{fig:model}
    \end{subfigure}
    \begin{subtable}[b]{0.4\linewidth}
    \begin{adjustbox}{width=\columnwidth,center}
    \footnotesize
    \begin{tabular}{lr}
    \hline
    \textbf{Inference Time (ms)} & \textbf{23.267} \\ \hline
    \textbf{Update Time (ms)}    & \textbf{65.242} \\ \hline
    \textbf{Warning Delay (ms)}  & \textbf{14.552} \\ \hline
    \textbf{Alarm Delay (s)}     & \textbf{1.670}  \\ \hline
                                 &                  \\ 
                                 &                  \\ \\
    \end{tabular}
    \end{adjustbox}
    \caption{\footnotesize{Latency of SiFall components.}}
    \label{t:overhead}
    \end{subtable}
    \caption{SiFall computation overhead.}
  \vspace{-0.4cm}
    
\end{figure}

We provide a quantitative analysis of the \systemnames{} computation overhead.
We utilize the Pytorch-OpCounter tool
to measure the computation cost in flops (floating-point operations per second) of FallNet 
and some representative CNN-based models used in other applications. 
As Figure~\ref{fig:model} depicts, FallNet falls in between the ultra-lightweight model (e.g., MobileNetV2) 
and the medium-weight models (eg., Densenet121 and AlexNet), which indicates FallNet is a relatively lightweight model.

We also measure the end to end latency of SiFall operation in our testbed, and summarize the result in Table~\ref{t:overhead}.
The average inference time and the model parameter update time are measured on a single NVIDIA 2080Ti GPU.
Once SiFall front-end detects a fall-like activity, it triggers a warning and waits for the FallNet at the back-end to generate the alarm for confirmed falls. We measure the warning delay (alarm delay) by averaging the time interval between the system warning time (alarm time) and the ground truth ending time of the fall-like activities (fall). The major delay of the system comes from the signal segmentation, where SiFall keeps monitoring the channel dynamics for ~1s.

\section{related work}

\textbf{RF-based fall detection.} 
Existing RF-based fall detection systems~\cite{wang2016wifall,palipana2018falldefi,tian2018rf,wang2016rt,hu2021defall,ruan2015tagfall} all assume repeatable human fall patterns and follow pre-defined fall templates in the feature space for detection. Most of them depend on manually segmented signal clips for inference.
Aryokee~\cite{tian2018rf} utilizes CNN to extract features of human fall as opposed to previous manual feature extraction approaches~\cite{wang2016wifall,palipana2018falldefi,tian2018rf,wang2016rt}. However, the samples are collected offline with the same length, the Aryokee model is not able to deal with varying length RF samples and the system cannot run in real-time. FallDefi~\cite{palipana2018falldefi} improves the performance by using the combined features of previous approaches and adopting more WiFi links for gathering RF signals. There are some general RF based human activity recognition systems including Witrack~\cite{adib20143d}, CARM~\cite{wang2015understanding} and HAR-SANet~\cite{chen2021rf} which treat fall as one of the ordinary human activities and can only capture few types of falls. To the best of our knowledge, RT-Fall is the most practical solution of real time fall detection, which however as suggested in our experimental evaluation cannot provide high accuracy with realistic falls occurring in practice. 
Although Defall~\cite{hu2021defall} claims real-time fall detection capability, it uses a human-like dummy to do the experiment to learn the fall template, and thus it can only detect simulated "hard fall", which is falling from a standstill position at a certain height, by its nature significantly limits its application in practice. 

\parahead{Other fall detection solutions}
Other than RF-based fall detection, there are CV-based fall detection approaches~\cite{de2017home,yu2017computer,galvao2017anomaly} that take optical measurements by camera or infrared sensors for analysis. Those solutions are often criticized for compromising human privacy. Wearable-based fall detection methods either require the user to carry the device~\cite{aguiar2014accelerometer,carletti2017smartphone} or wear the device~\cite{kaewkannate2016comparison} which are intrusive and thus not the most desired way for fall detection~\cite{ramachandran2020survey}. The acoustic-based~\cite{despotovic2022audio} method is limited by ambient noise. Sensor fusion-based fall detection~\cite{li2019bi, zhou2018fall} is believed to be more reliable as various sensors may complement each other in different situations, but generally leads to higher cost and deployment overhead. Among them, some works claim they detect falls based on anomaly detection~\cite{galvao2017anomaly, carletti2017smartphone}. A CV-based approach~\cite{galvao2017anomaly} collects a balanced dataset with fall and non-fall samples and use a supervised anomaly detection method. A wearable-based approach~\cite{carletti2017smartphone} learns a fixed boundary in feature space to separate daily activity and the anomaly fall, which cannot cope with unseen daily activities and falls.

\parahead{Deep learning based RF sensing}
Deep learning has recently been widely adopted to various wireless sensing applications, including physiological sensing~\cite{zhao2017learning}, food and liquid sensing~\cite{ha2020food}, gesture recognition~\cite{zheng2019zero}, body skeletons reconstructing~\cite{zhao2018rf,zhao2018through,liu2019real}, localization~\cite{ayyalasomayajula2020deep} and \textit{etc}~\cite{cai2020teaching,jiang2020towards,gong2021rf}.Those solutions cannot be directly applied to detecting human falls. Most of them do not support the neural network update during run time and often require extensive data collection and annotation to facilitate the model training.

\parahead{Anomaly detection}
While anomaly detection is well-studied in the literature, anomaly detection for high-dimensional data in real-time remains challenging~\cite{pang2021deep}. Traditional methods such as One-Class SVM~\cite{scholkopf2001estimating}, Kernel Density Estimation~\cite{parzen1962estimation} and Tree-based Isolation Forest~\cite{liu2008isolation} all fail to operate online due to unsatisfactory computational scalability and the curse of dimensionality. Thanks to the rapid development of deep learning technology, a lot of deep learning based anomaly detection methods have been proposed~\cite{xu2015learning,zong2018deep,wang2019effective} with similar frameworks that consist of three parts: feature extraction, feature representation learning, and end-to-end anomaly score learning. We design FallNet based on this skeleton and make it capable to run in real-time with unstructured input signal data, to fill the gap in the literature, as most deep learning based methods are capable to only structured datasets and lack real-time practices.

\parahead{Self-supervised learning}
Self-supervised learning techniques support learning representations from a large amount of unlabeled data and based on that representation to serve downstream classification tasks with a few labeled instances~\cite{jing2020self}. As an alternative solution to establish a representation of daily human activities, self-supervised learning still faces the challenge in the lack of labels for unforeseeable human fall types. From a different perspective, SiFall deals with the domain variations by building an anomaly detection neural network model and continuously evolving the model to represent high-level semantics of normal daily activities.

\section{Discussion \& Conclusion}
This paper proposes SiFall, a self-supervised incremental
learning human fall detection system. SiFall leverages Wi-Fi RF signals and is able to detect daily human falls in real time. Extensive experiment results demonstrate that SiFall achieves high accuracy in human fall detection and is resilient to varied human subjects, environment, and different types of falls. The design of SiFall makes an important contribution towards building practical and reliable RF-based fall detection systems. The current study is still limited in its lack of real fall samples, especially of elderly aged above 60. Since SiFall relies on wireless channel dynamics to catch human activities, it is currently limited to working with single room occupancy. We leave the exploration to the above two limitations to future work when developing SiFall into higher technology readiness levels.

\begin{acks}
We sincerely thank the shepherd and reviewers for their insightful comments and suggestions. We also thank all volunteers for their participation in our experiments. This research is supported by the National Research Foundation Singapore under its Industry Alignment Fund – Pre-positioning (IAF-PP) Funding Initiative, and Ministry of Education Singapore MOE AcRF Tier 2 MOE-T2EP20220-0004. Any opinions, findings and conclusions or recommendations expressed in this material are those of the authors and do not reflect the views of National Research Foundation Singapore and other funding agencies.
\end{acks}

\clearpage
\bibliographystyle{ACM-Reference-Format}
\bibliography{sample-base}
\appendix
\section{Instance Norm vs Batch Norm}
The equation of Instance Norm is the same as Batch Norm such that: 
\begin{equation}
  \mathrm{B} \mathrm{N}_{\gamma, \beta}\left(x\right) \equiv \gamma\left(\frac{x-\mu(x)}{\sigma(x)}\right)+\beta
\end{equation}
\begin{equation}
  \mathrm{I} \mathrm{N}_{\gamma, \beta}\left(x\right) \equiv \gamma\left(\frac{x-\mu(x)}{\sigma(x)}\right)+\beta
\end{equation}
where $\gamma, \beta$ are affine parameters learned from data; $\mu(x),\sigma(x)$  are the mean and standard deviation. The difference of the two norm just the way that how the statistical descriptors $\mu$ and $\sigma$ are obtained.

Given an input batch $x \in \mathbb{R}^{B  \times H \times W \times C}$, Batch Norm normalizes the mean and standard deviation for each individual feature channel to a whole batch:

\begin{small}
\begin{equation}
\mu_{c}(x)=\frac{1}{B H W} \sum_{n=1}^{B} \sum_{h=1}^{H} \sum_{w=1}^{W} x_{b c h w}
\end{equation}
\begin{equation}
\sigma_{c}(x)=\sqrt{\frac{1}{B H W}} \sum_{n=1}^{B} \sum_{h=1}^{H} \sum_{w=1}^{W}\left(x_{b c h w}-\mu_{c}(x)\right)^{2}+\epsilon   
\end{equation}
\end{small}
As Batch Norm uses mini-batch statistics during training phase and replace them with average mean and variance across batches during inference phase, which implicitly requires consistency distribution of training domain and inference domain. SiFall is an online detection system and the samples keep generated that might induce difference across different person and environments so that we use Instance Norm which normalize as per sample:
\begin{small}
\begin{equation}
\mu_{bc}(x)=\frac{1}{H W} \sum_{h=1}^{H} \sum_{w=1}^{W} x_{b c h w}
\end{equation}
\begin{equation}
\sigma_{bc}(x)=\sqrt{\frac{1}{ H W}} \sum_{h=1}^{H} \sum_{w=1}^{W}\left(x_{b c h w}-\mu_{bc}(x)\right)^{2}+\epsilon   
\end{equation}
\end{small}
 
\section{Convolution operation independent to the input size.}
The "convolution" operation in the neural network is different from the "convolution" in the signal processing domain. Indeed, the convolution operation is an element-wise multiplication and summation over a local region of the input tensor. The operation is repeated in sequential local regions until the whole tensor has been calculated.  

Each learnable filter $W$ in convolution operation with dimension $W \in \mathbb{R}^{k \times k}$, where $k$ denotes the kernel size, i.e., the size of the local region that calculates the multiplication. Let $X\in \mathbb{R}^{H_{in} \times W_{in} \times C}$ denote the input tensor. The convolution operation calculates output $Y$ such that:
\begin{equation*}
\small
Y_{p, q}=\sum_{n=1}^{C} \sum_{i, j \in \mathcal{N}_{k}} W_{i+\frac{k-1}{2}, j+\frac{k-1}{2}}^{\top} X_{p+i, q+j}^{n}
\end{equation*}

where $(p, q)$ denotes the location coordinate and

$\mathcal{N}_{k}=$ \allowbreak
$\left\{(i, j): i=\left\{-\frac{k-1}{2}, \ldots, \frac{k-1}{2}\right\}, j=\left\{-\frac{k-1}{2}, \ldots, \frac{k-1}{2}\right\}\right\}$ defines a local neighborhood. A convolution layer specify how the kernel sliding $i$ and $j$ through the input tensor by setting stride $s$ and how we want the input tensor be padded by setting $p$, as a result the output $ Y \in \mathbb{R}^{H_{out} \times W_{out} \times f}$ can be computed by
\[\left ( H_{out},W_{out} \right ) = \left ( \left \lfloor \frac{H_{in} + 2*p - k}{s} \right \rfloor + 1,\left \lfloor \frac{W_{in} + 2*p - k}{s} \right \rfloor + 1 \right )\], where $f$ is the number of learnable filters. The 'same padding' technique help choose proper $s$ and $p$ so that $H_{out} = H_{in}$, $W_{out} = W_{in}$. Consequently, convolution layer are able to adapt to the input tensor with arbitrary size.

\end{document}